\documentclass{article}
\usepackage{amsmath,amssymb,amsfonts}

\def\op#1{\mathop{{\it\fam0} #1}\limits}

\newcommand{\beq}{\begin{equation}}
\newcommand{\eeq}{\end{equation}}
\newcommand{\ben}{\begin{eqnarray}}
\newcommand{\een}{\end{eqnarray}}
\newcommand{\be}{\begin{eqnarray*}}
\newcommand{\ee}{\end{eqnarray*}}
\newcommand{\bea}{\begin{eqalph}}
\newcommand{\eea}{\end{eqalph}}

\newcommand{\di}{{\mathrm {dim}\,}}

\newcommand{\hm}{{\mathrm{Hom}\,}}
\newcommand{\pr}{{\mathrm{pr}\,}}
\newcommand{\dif}{{\mathrm {Diff}\,}}
\newcommand{\im}{{\mathrm {Im}\, }}
\newcommand{\llr}{\op\longleftarrow}

\newcommand{\al}{\alpha}
\newcommand{\bt}{\beta}
\newcommand{\dl}{\delta}
\newcommand{\la}{\lambda}
\newcommand{\La}{\Lambda}
\newcommand{\f}{\phi}
\newcommand{\vf}{\varphi}

\newcommand{\om}{\omega}

\newcommand{\m}{\mu}
\newcommand{\n}{\nu}
\newcommand{\g}{\gamma}

\newcommand{\vr}{\varrho}
\newcommand{\thh}{\theta}
\newcommand{\vt}{\vartheta}
\newcommand{\cG}{{\mathfrak g}}
\newcommand{\ve}{\varepsilon}
\newcommand{\up}{\upsilon}
\newcommand{\e}{\epsilon}
\newcommand{\ap}{\approx}
\newcommand{\rdr}{\stackrel{\leftarrow}{\dr}{}}

\newcommand{\bll}{\bullet}
\newcommand{\bbc}{{\bf b}}

\newcommand{\nw}[1]{[{#1}]}
\newcommand{\nm}[1]{|{#1}|}

\newcommand{\id}{{\mathrm{Id}\,}}

\newcommand{\si}{\sigma}
\newcommand{\Si}{\Sigma}

\newcommand{\lto}{{\leftarrow}}

\newcommand{\cO}{{\mathcal O}}
\newcommand{\cA}{{\mathcal A}}
\newcommand{\sC}{{\mathcal C}}
\newcommand{\cJ}{{\mathcal J}}

\newcommand{\gL}{{\mathfrak L}}

\newcommand{\gd}{{\mathfrak d}}

\newcommand{\gA}{{\mathfrak A}}

\newcommand{\cT}{{\mathcal T}}
\newcommand{\cP}{{\mathcal P}}
\newcommand{\cR}{{\mathcal R}}
\newcommand{\cL}{{\mathcal L}}
\newcommand{\cV}{{\mathcal V}}

\newcommand{\cE}{{\mathcal E}}

\newcommand{\cF}{{\mathcal F}}
\newcommand{\cC}{{\mathcal C}}

\newcommand{\cK}{{\mathcal K}}

\newcommand{\ccG}{{\mathcal G}}

\newcommand{\cS}{{\mathcal S}}

\newcommand{\bL}{{\mathbf L}}

\newcommand{\bE}{{\mathbf E}}

\newcommand{\bb}{{\mathbf 1}}

\newcommand{\w}{\wedge}

\newcommand{\wh}{\widehat}
\newcommand{\ol}{\overline}

\newcommand{\dr}{\partial}

\newcommand{\ar}{\op\longrightarrow}

\newcommand{\ot}{\otimes}

\newenvironment{eqalph}{\stepcounter{equation}
\setcounter{equationa}{\value{equation}} \setcounter{equation}{0}

\begin{eqnarray}}{\end{eqnarray}\setcounter{equation}{\value{equationa}}}

\newcounter{equationa}
\newcounter{remark}
\newcounter{example}
\newcounter{theorem}
\newcounter{proposition}
\newcounter{lemma}
\newcounter{corollary}
\newcounter{definition}
\def\theremark{\arabic{remark}}

\def\thedefinition{\arabic{theorem}}

\newenvironment{remark}{\refstepcounter{remark} \medskip {\bf Remark
\theremark.} }{ \medskip }
\newenvironment{example}{\refstepcounter{remark} \medskip {\bf
Example \theremark.} }{ \medskip }
\newenvironment{theorem}{\refstepcounter{theorem} \medskip{\bf
Theorem \thedefinition.}}{\medskip }

\newenvironment{lemma}{\refstepcounter{theorem} \medskip{\bf  Lemma
\thedefinition.}}{\medskip }
\newenvironment{corollary}{\refstepcounter{theorem} \medskip{\bf
Corollary \thedefinition.} }{\medskip }
\newenvironment{definition}{\refstepcounter{theorem} \medskip{\bf
Definition \thedefinition.} }{\medskip }

\newcommand{\mar}[1]{}

\begin{document}

\hbox{}

\begin{center}

{\Large\bf SUSY gauge theory on graded manifolds}

\bigskip

G. Sardanashvily, W. Wachowski

\medskip

Department of Theoretical Physics, Moscow State University,
Moscow, Russia

\bigskip

\end{center}

\begin{abstract}
Lagrangian classical field theory of even and odd fields is
adequately formulated in terms of fibre bundles and graded
manifolds. In particular, conventional Yang--Mills gauge theory is
theory of connections on smooth principal bundles, but its BRST
extension involves odd ghost fields an antifields on graded
manifolds. Here, we formulate Yang--Mills theory of
Grassmann-graded gauge fields associated to Lie superalgebras on
principal graded bundles. A problem lies in a geometric definition
of odd gauge fields. Our goal is Yang--Mills theory of graded
gauge fields and its BRST extension.
\end{abstract}

\section{Introduction}

Conventional Yang--Mills theory of classical gauge fields is
adequately formulated as Lagrangian theory of principal
connections on smooth principal bundles \cite{book09,book00}.
Here, we aim to develop Yang--Mills theory of Grassmann-graded
even and odd gauge fields associated to Lie superalgebras which
characterize various SUSY extensions of Standard Model. A key
point is the geometric description of odd gauge fields. A main
contradiction is that gauge fields are affine objects, whereas odd
fields are linear. A problem also lies in definition of odd fields
and their jets.

Even classical fields on an $n$-dimensional smooth manifold $X$
are adequately represented by sections of some fibre bundle $Y\to
X$, and their Lagrangian theory is comprehensively formulated in
terms of the variational bicomplex on jet manifolds $J^*Y$ of $Y$
\cite{book09,book13}. This formulation is based on the categorial
equivalence of projective $C^\infty(X)$-modules of finite ranks
and vector bundles over $X$ in accordance with the well-known
Serre--Swan theorem, generalized to an arbitrary manifold
\cite{book09,ren}.

At the same time, different geometric models of odd variables
either on graded manifolds or supermanifolds are discussed
\cite{cari03,cia95,franc,mont92,mont,sard13}. Both graded
manifolds and supermanifolds are phrased in terms of sheaves of
graded commutative algebras \cite{bart,book09,sard09}. However,
graded manifolds are characterized by sheaves on smooth manifolds,
while supermanifolds are constructed by gluing of sheaves on
supervector spaces. Treating odd fields on a smooth manifold $X$,
we follow the above mentioned Serre--Swan theorem extended to
graded manifolds (Theorem \ref{vv0}) \cite{sard13,SS}. It states
that, if a graded commutative $C^\infty(X)$-ring is generated by a
projective $C^\infty(X)$-module of finite rank, it is isomorphic
to a ring of graded functions on a graded manifold whose body is
$X$. In accordance with this theorem, we describe odd fields in
terms of graded manifolds \cite{book09,book13,sard13}.

Let us recall that a graded manifold is a locally-ringed space,
characterized by a smooth body manifold $Z$ and some structure
sheaf $\gA$ of Grassmann algebras on $Z$
\cite{bart,book09,sard09}. This sheaf is projected as $\si$
(\ref{cmp140}) onto a sheaf $C^\infty_Z$ of smooth real functions
on $Z$. Its sections form a graded commutative $C^\infty(Z)$-ring
$\cA$ of graded functions on a graded manifold $(Z,\gA)$. It is
called the structure ring of $(Z,\gA)$. The differential calculus
on a graded manifold is defined as the Chevalley--Eilenberg
differential calculus on its structure ring (Section 3). However,
it is not a particular noncommutative differential calculus
because the Leibniz rule (\ref{ws10}) of derivations of a graded
commutative ring differs from that of a noncommutative ring
\cite{book13}.

It is important for our consideration that, by virtue of the
well-known Batchelor theorem (Theorem \ref{lmp1a}), there exists a
vector bundle $E\to Z$ with a typical fibre $V$ such that the
structure sheaf $\gA$ of $(Z,\gA)$ is isomorphic to a sheaf
$\gA_E$ of germs of sections of the exterior bundle $\w E^*$ of
the dual $E^*$ of $E$ whose typical fibre is the Grassmann algebra
$\w V^*$ \cite{bart,batch1}. This Batchelor's isomorphism is not
canonical. In field models, it however is fixed from the
beginning. Therefore, we restrict our consideration to graded
manifolds $(Z,\gA_E)$, called the simple graded manifolds,
modelled over vector bundles $E\to Z$.

Let us note that a smooth manifold $Z$ itself can be treated as a
trivial graded manifold $(Z,C^\infty_Z)$ whose structure ring of
graded functions is reduced to a ring $C^\infty(Z)$ of smooth real
functions on $Z$ (Remark \ref{triv}). Accordingly, a fibre bundle
$Y\to X$ in a theory of even classical fields can be regarded as a
graded bundle of trivial graded manifolds $(Y, C^\infty_Y) \to (X,
C^\infty_X)$ (Remark \ref{su20}). It follows that, in a general
setting, one can define a configuration space of theory of even
and odd classical fields as being a graded bundle of graded
manifolds $(Y, \gA) \to (X, \gA')$, where $Y\to X$ is a smooth
bundle (Definition \ref{su23}).

Considering even and odd fields on a smooth manifold $X$, we
however restrict our analysis to graded bundles
\mar{su11}\beq
(Y, \gA) \to (X, C^\infty_X). \label{su11}
\eeq
over a trivial graded manifold $(X, C^\infty_X)$ (Definition
\ref{su22}). We call it the graded bundle over a smooth manifold
$X$. It is denoted by $(X,Y,\gA)$. If $Y\to X$ is a vector bundle,
this is a particular case of graded vector bundles in
\cite{hern,mont} whose base is a trivial graded manifold.
Accordingly, $(X,Y,C^\infty_Y)$ denotes a fibre bundle $Y\to X$.

As was mentioned above, Lagrangian field theory on a fibre bundle
$Y\to X$ is formulated in terms of the variational bicomplex on
jet manifolds $J^*Y$ of $Y$. These are fibre bundles over $X$ and,
therefore, they can be regarded as trivial graded bundles $(X,
J^kY, C^\infty_{J^kY})$. Then let us define their partners in the
case of the non-trivial graded manifold (\ref{su11}) as follows
(Section 6).

Let $(Y, \gA_F)$ (\ref{su11}) be a simple graded manifold modelled
over a vector bundle $F\to Y$ where $Y\to X$ is a fibre bundle. We
have a composite bundle $F\to Y\to X$. Let us consider a $k$-order
jet manifold $J^kF$ of $F\to X$. This is a composite bundle
$J^kF\to J^kY\to X$ which is a vector bundle $J^kF\to J^kY$ over a
$k$-order jet manifold $J^kY$ of $Y\to X$. Then one can consider a
simple graded manifold $(J^kY, \gA_{J^kF})$ which also is a graded
bundle $(X,J^kY, \gA_{J^kF})$ over a smooth manifold $X$. We agree
to call it the graded $k$-order jet manifold of a graded bundle
$(X,Y, \gA_F)$ (Definition \ref{su32}). Obviously, the above
mentioned trivial graded jet manifold $(X, J^kY, C^\infty_{J^kY})$
of $(Y, \gA_{F=Y\times\{0\}}=C^\infty_Y)$ is a trivial graded
bundle \cite{book09,sard13}.

This definition differs from that of graded jet bundles in
\cite{hern,mont}, but it reproduces the heuristic notion of jets
of odd ghosts in field-antifield BRST theory \cite{barn,bran01}.

Furthermore, the inverse sequence of jet manifolds
\mar{j1}\beq
Y\op\longleftarrow^\pi J^1Y \longleftarrow \cdots J^{r-1}Y
\op\longleftarrow^{\pi^r_{r-1}} J^rY\longleftarrow\cdots,
\label{j1}
\eeq
yields the inverse sequence of graded jet manifolds
\mar{su14}\beq
(Y,\cA_F)\op\longleftarrow (J^1Y,\gA_{J^1F}) \longleftarrow \cdots
(J^{r-1}Y, \gA_{J^{r-1}F}) \op\longleftarrow (J^rY,
\gA_{J^rF})\longleftarrow\cdots. \label{su14}
\eeq
One can think on its inverse limit $(J^\infty Y, \cA_{J^\infty
F})$ as the graded infinite order jet manifold  whose body is an
infinite order jet manifold $J^\infty Y$ and whose structure sheaf
$\cA_{J^\infty F}$ is a sheaf of germs of graded functions on
graded manifolds $(J^*Y,\gA_{J^*F})$. However $(J^\infty Y,
\cA_{J^\infty F})$ fails to be a graded manifold in a strict sense
because the inverse image $J^\infty Y$ of the sequence (\ref{j1})
is a Fr\'eche manifold, but not the smooth one.

Classical Lagrangian theory of even and odd fields is
comprehensively formulated in terms of a Grassmann-graded
variational bicomplex on a graded infinite order jet manifold
$(J^\infty Y, \cA_{J^\infty F})$ (Section 7) \cite{book09,sard13}.

However, quantization of Lagrangian field theory essentially
depends on its degeneracy characterized by a set of non-trivial
reducible Noether identities \cite{barn,lmp08,gom}. These Noether
identities can obey first-stage Noether identities, which in turn
are subject to the second-stage ones, and so on. Thus, there is a
hierarchy of Noether and higher-stage Noether identities
\cite{lmp08,book09}. Bearing in mind gauge theory, we here
restrict our consideration to irreducible Lagrangian systems whose
Noether identities are independent (Section 8). The inverse second
Noether theorem (Theorem \ref{w35}) associates to them the gauge
operator (\ref{w33}) which is a non-trivial gauge symmetry of an
original Lagrangian. Its nilpotent extension is the BRST operator
(\ref{w109}). If the BRST operator exists, it provides the BRST
extension of original Lagrangian field theory by Grassmann-graded
ghosts and antifields \cite{jmp09,book09}. This BRST extension is
a first step towards BV quantization of classical field theory in
terms of functional integrals \cite{barn,gom}.

As was mentioned above, classical gauge theory is formulated as
Lagrangian theory of principal connections on smooth principal
bundles (Section 9). Similarly, we develop SUSY gauge theory in
the framework of Grassmann-graded Lagrangian formalism on
principal graded bundles (Section 10). A key point is that we
consider simple principal graded bundles (Definition \ref{su55})
subject to an action of an even Lie group, but not the whole
graded Lie one. In this case, odd gauge potentials in comparison
with the even ones are linear, but not affine objects (see Remark
\ref{su92}). At the same time, they admit the affine gauge
transformations (\ref{su110}) parameterised by ghosts. Our goal is
Yang--Mills theory of graded gauge fields and its BRST extension
(Section 11).

\section{Grassmann-graded algebraic calculus}

Throughout this work, by the Grassmann gradation is meant $\mathbb
Z_2$-gradation, and a Grassmann graded structure simply is called
the graded structure if there is no danger of confusion.
Hereafter, the symbol $\nw .$ stands for the Grassmann parity. Let
us summarize the relevant notions of the Grassmann-graded
algebraic calculus \cite{bart,book09,sard09}.

Let $\cK$ be a commutative ring. A $\cK$-module $Q$ is called
graded if it is endowed with a grading automorphism $\g$,
$\g^2=\id$. A graded module falls into a direct sum of modules
$Q=Q_0 \oplus Q_1$ such that
\be
\g(q)=(-1)^{[q]}q, \qquad q\in Q_{[q]}.
\ee
One calls $Q_0$ and $Q_1$ the even and odd parts of $Q$,
respectively. A graded $\cK$-module is said to be free if it has a
basis composed by graded-homogeneous elements.

In particular, by a real graded vector space $B=B_0\oplus B_1$ is
meant a graded $\mathbb R$-module. A real graded vector space is
said to be $(n,m)$-dimensional if $B_0=\mathbb R^n$ and
$B_1=\mathbb R^m$.

A $\cK$-algebra $\cA$ is called graded if it is a graded
$\cK$-module such that
\be
[aa']=([a]+[a']){\rm mod}\,2,
\ee
where $a$ and $a'$ are graded-homogeneous elements of $\cA$. Its
even part $\cA_0$ is a subalgebra of $\cA$, and the odd one
$\cA_1$ is an $\cA_0$-module. If $\cA$ is a graded ring, then
$[\bb]=0$.

Given a graded algebra $\cA$, a left graded $\cA$-module $Q$ is
defined as a left $\cA$-module where
\be
[aq]=([a]+[q]){\rm mod}\,2.
\ee
Similarly, right graded $\cA$-modules are treated.

A graded algebra $\cA$ is called graded commutative if
\be
aa'=(-1)^{[a][a']}a'a.
\ee
In this case, a graded $\cA$-module $Q$ is provided with a graded
$\cA$-bimodule structure by letting
\be
qa = (-1)^{[a][q]}aq, \qquad a\in\cA, \qquad q\in Q.
\ee

Given a graded commutative ring $\cA$, the following are standard
constructions of new graded modules from old ones.

$\bullet$ The direct sum of graded modules and a graded factor
module are defined just as those of modules over a commutative
ring.

$\bullet$ The tensor product $P\ot Q$ of graded $\cA$-modules $P$
and $Q$ is their tensor product as $\cA$-modules such that
\be
&& [p\ot q]=[p]+ [q], \qquad p\in P, \qquad q\in Q, \\
&&  ap\ot q=(-1)^{[p][a]}pa\ot q= (-1)^{[p][a]}p\ot aq,  \qquad a\in\cA.
\ee
In particular, the tensor algebra $\ot P$ of a graded $\cA$-module
$P$ is defined just as that of a module over a commutative ring.
Its quotient $\w P$ with respect to the ideal generated by
elements
\be
p\ot p' + (-1)^{[p][p']}p'\ot p, \qquad p,p'\in P,
\ee
is the bigraded exterior algebra of a graded module $P$ with
respect to the graded exterior product
\be
p\w p' =- (-1)^{[p][p']}p'\w p.
\ee

$\bullet$ A morphism $\Phi:P\to Q$ of graded $\cA$-modules seen as
additive groups is said to be an even graded morphism (resp. an
odd graded morphism) if $\Phi$ preserves (resp. changes) the
Grassmann parity of all graded-homogeneous elements of $P$ and if
the relations
\be
\Phi(ap)=(-1)^{[\Phi][a]}a\Phi(p), \qquad p\in P, \qquad a\in\cA,
\ee
hold. A morphism $\Phi:P\to Q$ of graded $\cA$-modules as additive
groups is called a graded $\cA$-module morphism if it is
represented by a sum of even and odd graded morphisms. A set
$\hm_\cA(P,Q)$ of graded morphisms of a graded $\cA$-module $P$ to
a graded $\cA$-module $Q$ is naturally a graded $\cA$-module. A
graded $\cA$-module $P^*=\hm_\cA(P,\cA)$ is called the dual of a
graded $\cA$-module $P$.

Let $V$ be a real vector space, and let $\La=\w V$ be its exterior
algebra endowed with the Grassmann gradation
\mar{+66}\beq
\La=\La_0\oplus \La_1, \qquad \La_0=\mathbb R\op\bigoplus_{k=1}
\op\w^{2k} V, \qquad \La_1=\op\bigoplus_{k=1} \op\w^{2k-1} V.
\label{+66}
\eeq
It is a real graded commutative ring, called the Grassmann
algebra. A Grassmann algebra, seen as an additive group, admits
the decomposition
\mar{+11}\beq
\La=\mathbb R\oplus R =\mathbb R\oplus R_0\oplus R_1=\mathbb R
\oplus (\La_1)^2 \oplus \La_1, \label{+11}
\eeq
where $R$ is the ideal of nilpotents of $\La$. The corresponding
epimorphism $\si:\La\to\mathbb R$ is called the body map.

Note that there is a different definition of a Grassmann algebra
which is equivalent to the above mentioned one only in the case of
infinite-dimensional vector spaces $V$ \cite{jad}. Hereafter, we
restrict our consideration to Grassmann algebras of finite rank
when $V=\mathbb R^N$. Given a basis $\{c^i\}$ for $V$, elements of
the Grassmann algebra $\La$ (\ref{+66}) take a form
\mar{z784}\beq
a=\op\sum_{k=0,1,\ldots} \op\sum_{(i_1\cdots i_k)}a_{i_1\cdots
i_k}c^{i_1}\cdots c^{i_k}, \label{z784}
\eeq
where the second sum runs through all the tuples $(i_1\cdots i_k)$
such that no two of them are permutations of each other.

A graded algebra $\cG$ is called a Lie superalgebra if its product
$[.,.]$, called the Lie superbracket, obeys the relations
\be
&& [\ve,\ve']=-(-1)^{[\ve][\ve']}[\ve',\ve],\\
&& (-1)^{[\ve][\ve'']}[\ve,[\ve',\ve'']]
+(-1)^{[\ve'][\ve]}[\ve',[\ve'',\ve]] +
(-1)^{[\ve''][\ve']}[\ve'',[\ve,\ve']] =0.
\ee
Being decomposed in even and odd parts $\cG=\cG_0\oplus \cG_1$, a
Lie superalgebra $\cG$ obeys the relations
\be
[{\cG_0},{\cG_0}]\subset \cG_0, \qquad [{\cG_0},{\cG_1}]\subset
\cG_1, \qquad [{\cG_1},{\cG_1}]\subset \cG_1.
\ee
In particular, an even part $\cG_0$ of a Lie superalgebra $\cG$ is
a Lie algebra.

\section{Grassmann-graded differential calculus}

Linear differential operators on graded modules over a graded
commutative ring are defined similarly to those in commutative
geometry \cite{book09,sard09,book12}.

Let $\cK$ be a commutative ring and $\cA$ a graded commutative
$\cK$-ring. Let $P$ and $Q$ be graded $\cA$-modules. A
$\cK$-module $\hm_\cK (P,Q)$ of graded $\cK$-module homomorphisms
$\Phi:P\to Q$ can be endowed with the two graded $\cA$-module
structures
\mar{ws11}\beq
(a\Phi)(p)= a\Phi(p),  \qquad  (\Phi\bll a)(p) = \Phi (a p),\qquad
a\in \cA, \quad p\in P, \label{ws11}
\eeq
called $\cA$- and $\cA^\bll$-module structures, respectively. Let
us put
\mar{ws12}\beq
\dl_a\Phi= a\Phi -(-1)^{[a][\Phi]}\Phi\bll a, \qquad a\in\cA.
\label{ws12}
\eeq
An element $\Delta\in\hm_\cK(P,Q)$ is said to be a $Q$-valued
graded differential operator of order $s$ on $P$ if
$\dl_{a_0}\circ\cdots\circ\dl_{a_s}\Delta=0$ for any tuple of
$s+1$ elements $a_0,\ldots,a_s$ of $\cA$. A set $\dif_s(P,Q)$ of
these operators inherits the graded module structures
(\ref{ws11}).

In particular, zero order graded differential operators coincide
with graded $\cA$-module morphisms $P\to Q$. A first order graded
differential operator $\Delta$ satisfies a relation
\be
&& \dl_a\circ\dl_b\,\Delta(p)=
ab\Delta(p)- (-1)^{([b]+[\Delta])[a]}b\Delta(ap)-
(-1)^{[b][\Delta]}a\Delta(bp)+\\
&& \qquad (-1)^{[b][\Delta]+([\Delta]+[b])[a]}
=0, \qquad a,b\in\cA, \quad p\in P.
\ee

For instance, let $P=\cA$. Any zero order $Q$-valued graded
differential operator $\Delta$ on $\cA$ is defined by its value
$\Delta(\bb)$. Then there is a graded $\cA$-module isomorphism
\be
\dif_0(\cA,Q)=Q, \qquad Q\ni q\to \Delta_q\in \dif_0(\cA,Q),
\ee
where $\Delta_q$ is given by the equality $\Delta_q(\bb)=q$. A
first order $Q$-valued graded differential operator $\Delta$ on
$\cA$ fulfils a condition
\be
\Delta(ab)= \Delta(a)b+ (-1)^{[a][\Delta]}a\Delta(b)
-(-1)^{([b]+[a])[\Delta]} ab \Delta(\bb), \qquad  a,b\in\cA.
\ee
It is called the $Q$-valued graded derivation of $\cA$ if
$\Delta(\bb)=0$, i.e., a graded Leibniz rule
\mar{ws10}\beq
\Delta(ab) = \Delta(a)b + (-1)^{[a][\Delta]}a\Delta(b), \quad
a,b\in \cA, \label{ws10}
\eeq
holds. One then observes that any first order graded differential
operator on $\cA$ falls into a sum
\be
\Delta(a)= \Delta(\bb)a +[\Delta(a)-\Delta(\bb)a]
\ee
of a zero order graded differential operator $\Delta(\bb)a$ and a
graded derivation $\Delta(a)-\Delta(\bb)a$. If $\dr$ is a graded
derivation of $\cA$, then $a\dr$ is so for any $a\in \cA$. Hence,
graded derivations of $\cA$ constitute a graded $\cA$-module
$\gd(\cA,Q)$, called the graded derivation module. If $Q=\cA$, a
graded derivation module $\gd\cA$ also is a Lie superalgebra over
a commutative ring $\cK$ with respect to a superbracket
\mar{ws14}\beq
[u,u']=u\circ u' - (-1)^{[u][u']}u'\circ u, \qquad u,u'\in \cA.
\label{ws14}
\eeq

Since $\gd\cA$ is a Lie $\cK$-superalgebra, let us consider the
Chevalley--Eilenberg complex $C^*[\gd\cA;\cA]$ where a graded
commutative ring $\cA$ is a regarded as a $\gd\cA$-module
\cite{fuks,book09,book12}. It is the complex
\mar{ws85}\beq
0\to \cK\to \cA\ar^d C^1[\gd\cA;\cA]\ar^d \cdots
C^k[\gd\cA;\cA]\ar^d\cdots \label{ws85}
\eeq
where
\be
C^k[\gd\cA;\cA]=\hm_\cK(\op\w^k \gd\cA,\cA)
\ee
are $\gd\cA$-modules of $\cK$-linear graded morphisms of the
graded exterior products $\op\w^k \gd\cA$ of a graded $\cK$-module
$\gd\cA$ to $\cA$. Let us bring homogeneous elements of $\op\w^k
\gd\cA$ into the form
\be
\ve_1\w\cdots\ve_r\w\e_{r+1}\w\cdots\w \e_k, \qquad
\ve_i\in\gd\cA_0, \quad \e_j\in\gd\cA_1.
\ee
Then the Chevalley--Eilenberg coboundary operator $d$ of the
complex (\ref{ws85}) is given by the expression
\mar{ws86}\ben
&& dc(\ve_1\w\cdots\w\ve_r\w\e_1\w\cdots\w\e_s)=
\label{ws86}\\
&&\op\sum_{i=1}^r (-1)^{i-1}\ve_i
c(\ve_1\w\cdots\wh\ve_i\cdots\w\ve_r\w\e_1\w\cdots\e_s)+
\nonumber \\
&& \op\sum_{j=1}^s (-1)^r\ve_i
c(\ve_1\w\cdots\w\ve_r\w\e_1\w\cdots\wh\e_j\cdots\w\e_s)
+\nonumber\\
&& \op\sum_{1\leq i<j\leq r} (-1)^{i+j}
c([\ve_i,\ve_j]\w\ve_1\w\cdots\wh\ve_i\cdots\wh\ve_j
\cdots\w\ve_r\w\e_1\w\cdots\w\e_s)+\nonumber\\
&&\op\sum_{1\leq i<j\leq s} c([\e_i,\e_j]\w\ve_1\w\cdots\w
\ve_r\w\e_1\w\cdots
\wh\e_i\cdots\wh\e_j\cdots\w\e_s)+\nonumber\\
&& \op\sum_{1\leq i<r,1\leq j\leq s} (-1)^{i+r+1}
c([\ve_i,\e_j]\w\ve_1\w\cdots\wh\ve_i\cdots\w\ve_r\w
\e_1\w\cdots\wh\e_j\cdots\w\e_s),\nonumber
\een
where the caret $\,\wh{}\,$ denotes omission.

It is easily justified that the complex (\ref{ws85}) contains a
subcomplex $\cO^*[\gd\cA]$ of $\cA$-linear graded morphisms. The
$\mathbb N$-graded module $\cO^*[\gd\cA]$ is provided with the
structure of a bigraded $\cA$-algebra with respect to the graded
exterior product
\mar{ws103'}\ben
&& \f\w\f'(u_1,...,u_{r+s})= \op\sum_{i_1<\cdots<i_r;j_1<\cdots<j_s} {\rm
Sgn}^{i_1\cdots i_rj_1\cdots j_s}_{1\cdots r+s} \f(u_{i_1},\ldots,
u_{i_r}) \f'(u_{j_1},\ldots,u_{j_s}), \label{ws103'} \\
&& \f\in \cO^r[\gd\cA], \qquad \f'\in \cO^s[\gd\cA], \qquad u_k\in \gd\cA,
\nonumber
\een
where $u_1,\ldots, u_{r+s}$ are graded-homogeneous elements of
$\gd\cA$ and
\be
u_1\w\cdots \w u_{r+s}= {\rm Sgn}^{i_1\cdots i_rj_1\cdots
j_s}_{1\cdots r+s} u_{i_1}\w\cdots\w u_{i_r}\w u_{j_1}\w\cdots\w
u_{j_s}.
\ee
The graded Chevalley--Eilenberg coboundary operator $d$
(\ref{ws86}) and the graded exterior product $\w$ (\ref{ws103'})
bring $\cO^*[\gd\cA]$ into a differential bigraded algebra whose
elements obey relations
\mar{ws45}\beq
\f\w \f'=(-1)^{|\f||\f'|+[\f][\f']}\f'\w\f, \qquad  d(\f\w\f')=
d\f\w\f' +(-1)^{|\f|}\f\w d\f'. \label{ws45}
\eeq
It is called the graded Chevalley--Eilenberg differential calculus
over a graded commutative $\cK$-ring $\cA$. In particular, we have
\mar{ws47}\beq
\cO^1[\gd\cA]=\hm_\cA(\gd\cA,\cA)=\gd\cA^*. \label{ws47}
\eeq
One can extend this duality relation to the graded interior
product of $u\in\gd\cA$ with any element $\f\in \cO^*[\gd\cA]$ by
the rules
\mar{ws46}\ben
&& u\rfloor(bda) =(-1)^{[u][b]}bu(a),\qquad a,b \in\cA, \nonumber\\
&& u\rfloor(\f\w\f')=
(u\rfloor\f)\w\f'+(-1)^{|\f|+[\f][u]}\f\w(u\rfloor\f').
\label{ws46}
\een
As a consequence, any graded derivation $u\in\gd\cA$ of $\cA$
yields a derivation
\mar{+117}\ben
&& \bL_u\f= u\rfloor d\f + d(u\rfloor\f), \qquad \f\in\cO^*, \qquad
u\in\gd\cA, \label{+117} \\
&& \bL_u(\f\w\f')=\bL_u(\f)\w\f' + (-1)^{[u][\f]}\f\w\bL_u(\f'), \nonumber
\een
called the graded Lie derivative of the differential bigraded
algebra $\cO^*[\gd\cA]$.

The minimal graded Chevalley--Eilenberg differential calculus
$\cO^*\cA\subset \cO^*[\gd\cA]$  over a graded commutative ring
$\cA$ consists of the monomials $a_0da_1\w\cdots\w da_k, \qquad
a_i\in\cA$. The corresponding complex
\mar{t100}\beq
0\to\cK\ar \cA\ar^d\cO^1\cA\ar^d \cdots  \cO^k\cA\ar^d \cdots
\label{t100}
\eeq
is called the bigraded de Rham complex of a graded commutative
$\cK$-ring $\cA$.

\section{Graded manifolds}

A graded manifold of dimension $(n,m)$ is defined as a
local-ringed space $(Z,\gA)$ where $Z$ is an $n$-dimensional
smooth manifold $Z$ and $\gA=\gA_0\oplus\gA_1$ is a sheaf of
Grassmann algebras $\Lambda$ of rank $m$ such that
\cite{bart,book09,sard09}:

$\bullet$ there is the exact sequence of sheaves
\mar{cmp140}\beq
0\to \cR \to\gA \op\to^\si C^\infty_Z\to 0, \qquad
\cR=\gA_1+(\gA_1)^2,\label{cmp140}
\eeq
where $\si$ is a body epimorphism onto a sheaf $C^\infty_Z$ of
smooth real functions on $Z$;

$\bullet$ $\cR/\cR^2$ is a locally free sheaf of
$C^\infty_Z$-modules of finite rank (with respect to pointwise
operations), and the sheaf $\gA$ is locally isomorphic to the
exterior product $\w_{C^\infty_Z}(\cR/\cR^2)$.

The sheaf $\gA$ is called a structure sheaf of a graded manifold
$(Z,\gA)$, and a manifold $Z$ is said to be the body of $(Z,\gA)$.
Sections of the sheaf $\gA$ are called graded functions on a
graded manifold $(Z,\gA)$. They make up a graded commutative
$C^\infty(Z)$-ring $\gA(Z)$ called the structure ring of
$(Z,\gA)$.

A graded manifold $(Z,\gA)$ possesses the following local
structure. Given a point $z\in Z$, there exists its open
neighborhood $U$, called a splitting domain, such that
\be
\gA(U)= C^\infty(U)\ot\w\mathbb R^m.
\ee
This means that the restriction $\gA|_U$ of the structure sheaf
$\gA$ to $U$ is isomorphic to the sheaf $C^\infty_U\ot\w\mathbb
R^m$ of sections of some exterior bundle $U\times \w\mathbb R^m\to
U$. The well-known Batchelor theorem \cite{bart,batch1} states
that such a structure of a graded manifold is global as follows.

\begin{theorem} \label{lmp1a} \mar{lmp1a}
Let $(Z,\gA)$ be a graded manifold. There exists a vector bundle
$E\to Z$ with an $m$-dimensional typical fibre $V$ such that the
structure sheaf $\gA$ of $(Z,\gA)$ is isomorphic to the structure
sheaf $\gA_E$ of germs of sections of the exterior bundle $\w
E^*\to Z$, whose typical fibre is a Grassmann algebra $\w V^*$.
\end{theorem}

Combining Batchelor Theorem \ref{lmp1a} and the classical
Serre--Swan theorem leads to the following Serre--Swan theorem for
graded manifolds \cite{jmp05a,SS}.

\begin{theorem} \label{vv0} \mar{vv0}
Let $Z$ be a smooth manifold. A graded commutative
$C^\infty(Z)$-algebra $\cA$ is isomorphic to the structure ring of
a graded manifold with a body $Z$ iff it is the exterior algebra
of some projective $C^\infty(Z)$-module of finite rank.
\end{theorem}

It should be emphasized that Batchelor's isomorphism in Theorem
\ref{lmp1a} fails to be canonical. We agree to call $(Z,\gA_E)$ in
Theorem \ref{lmp1a} the simple graded manifold modelled over a
vector bundle $E\to Z$, named its characteristic vector bundle.
Accordingly, a structure ring $\gA_E(Z)$ of a simple graded
manifold $(Z,\gA_E)$ is a structure module
\mar{33f1}\beq
\cA_E=\gA_E(Z)=\w E^*(Z) \label{33f1}
\eeq
of sections of the exterior bundle $\w E^*$.

\begin{remark} \label{triv} \mar{triv}
Hereafter, it is convenient to treat a local-ringed space
\mar{trv}\beq
(Z,\gA_0=C_Z^\infty) \label{trv}
\eeq
as a trivial graded manifold. It is a simple graded manifold whose
characteristic bundle is $E=Z\times\{0\}$. Its structure module is
a ring $C^\infty(Z)$ of smooth real functions on $Z$.
\end{remark}

Given a simple graded manifold $(Z,\gA_E)$, every trivialization
chart $(U; z^A,q^a)$ of a vector bundle $E\to Z$ yields a
splitting domain $(U; z^A,c^a)$ of $(Z,\gA_E)$ where $\{c^a\}$ is
the corresponding local fibre basis for $E^*\to X$, i.e., $c^a$
are locally constant sections of $E^*\to X$ such that $q_b\circ
c^a=\dl^a_b$. Graded functions on such a chart are $\La$-valued
functions
\mar{z785}\beq
f=\op\sum_{k=0}^m \frac1{k!}f_{a_1\ldots a_k}(z)c^{a_1}\cdots
c^{a_k}, \label{z785}
\eeq
where $f_{a_1\cdots a_k}(z)$ are smooth functions on $U$. One
calls $\{z^A,c^a\}$ the local basis for a graded manifold
$(Z,\gA_E)$. Transition functions $q'^a=\rho^a_b(z^A)q^b$ of
bundle coordinates on $E\to Z$ induce the corresponding
transformation
\mar{+6}\beq
c'^a=\rho^a_b(z^A)c^b \label{+6}
\eeq
of the associated local basis for a graded manifold $(Z,\gA_E)$
and the according coordinate transformation law of graded
functions (\ref{z785}).

Note that, in general, automorphisms of a graded manifold take a
form $c'^a=\rho^a(z^A,c^b)$, where $\rho^a(z^A,c^b)$ are local
graded functions. Considering a simple graded manifold
$(Z,\gA_E)$, one restricts the class of graded manifold
transformations to the linear ones (\ref{+6}), compatible with
given Batchelor's isomorphism.

Let us consider the graded derivation module $\gd\gA(Z)$ of a real
graded commutative ring $\gA(Z)$. It is a real Lie superalgebra
relative to the superbracket (\ref{ws14}). Its elements are called
the graded vector fields on a graded manifold $(Z,\gA)$. A key
point is that graded vector fields $u\in\gd\cA_E$ on a simple
graded manifold $(Z,\gA_E)$ are represented by sections of some
vector bundle as follows \cite{book09}.

Due to the canonical splitting $VE= E\times E$, the vertical
tangent bundle $VE$ of $E\to Z$ can be provided with the fibre
bases $\{\dr/\dr c^a\}$, which are the duals of the bases
$\{c^a\}$. Then graded vector fields on a splitting domain
$(U;z^A,c^a)$ of $(Z,\gA_E)$ read
\mar{hn14}\beq
u= u^A\dr_A + u^a\frac{\dr}{\dr c^a}, \label{hn14}
\eeq
where $u^A, u^a$ are local graded functions on $U$. In particular,
\be
\frac{\dr}{\dr c^a}\circ\frac{\dr}{\dr c^b} =-\frac{\dr}{\dr
c^b}\circ\frac{\dr}{\dr c^a}, \qquad \dr_A\circ\frac{\dr}{\dr
c^a}=\frac{\dr}{\dr c^a}\circ \dr_A.
\ee
The graded derivations (\ref{hn14}) act on graded functions
$f\in\gA_E(U)$ (\ref{z785}) by the rule
\mar{cmp50a}\beq
u(f_{a\ldots b}c^a\cdots c^b)=u^A\dr_A(f_{a\ldots b})c^a\cdots c^b
+u^k f_{a\ldots b}\frac{\dr}{\dr c^k}\rfloor (c^a\cdots c^b).
\label{cmp50a}
\eeq
This rule implies the corresponding coordinate transformation law
\be
u'^A =u^A, \qquad u'^a=\rho^a_ju^j +u^A\dr_A(\rho^a_j)c^j
\ee
of graded vector fields. It follows that graded vector fields
(\ref{hn14}) are sections of the following vector bundle $\cV_E\to
Z$. This vector bundle is locally isomorphic to the vector bundle
\be
\cV_E|_U\approx\w E^*\op\ot_Z(E\op\oplus_Z TZ)|_U,
\ee
and it is characterized by an atlas of bundle coordinates
$(z^A,z^A_{a_1\ldots a_k},v^i_{b_1\ldots b_k})$, $k=0,\ldots,m$,
possessing transition functions
\be
&& z'^A_{i_1\ldots
i_k}=\rho^{-1}{}_{i_1}^{a_1}\cdots
\rho^{-1}{}_{i_k}^{a_k} z^A_{a_1\ldots a_k}, \\
&& v'^i_{j_1\ldots j_k}=\rho^{-1}{}_{j_1}^{b_1}\cdots
\rho^{-1}{}_{j_k}^{b_k}\left[\rho^i_jv^j_{b_1\ldots b_k}+
\frac{k!}{(k-1)!} z^A_{b_1\ldots b_{k-1}}\dr_A\rho^i_{b_k}\right].
\ee

Given a structure ring $\cA_E$ of graded functions on a simple
graded manifold $(Z,\gA_E)$ and the real Lie superalgebra
$\gd\cA_E$ of its graded derivations, let us consider the graded
Chevalley--Eilenberg differential calculus
\mar{33f21}\beq
\cS^*[E;Z]=\cO^*[\gd\cA_E] \label{33f21}
\eeq
over $\cA_E$. Since the graded derivation module $\gd\cA_E$ is
isomorphic to the structure module of sections of a vector bundle
$\cV_E\to Z$, elements of $\cS^*[E;Z]$ are represented by sections
of the exterior bundle $\w\ol\cV_E$ of the $\cA_E$-dual
$\ol\cV_E\to Z$ of $\cV_E$. With respect to the dual fibre bases
$\{dz^A\}$ for $T^*Z$ and $\{dc^b\}$ for $E^*$, sections of
$\ol\cV_E$ take a coordinate form
\be
\f=\f_A dz^A + \f_adc^a,
\ee
together with transition functions
\be
\f'_a=\rho^{-1}{}_a^b\f_b, \qquad \f'_A=\f_A
+\rho^{-1}{}_a^b\dr_A(\rho^a_j)\f_bc^j.
\ee
The duality isomorphism $\cS^1[E;Z]=\gd\cA_E^*$ (\ref{ws47}) is
given by the graded interior product
\be
u\rfloor \f=u^A\f_A + (-1)^{\nw{\f_a}}u^a\f_a.
\ee
Elements of $\cS^*[E;Z]$ are called graded exterior forms on a
graded manifold $(Z,\gA_E)$.

Seen as an $\cA_E$-algebra, the differential bigraded algebra
$\cS^*[E;Z]$ (\ref{33f21}) on a splitting domain $(U;z^A,c^a)$ is
locally generated by graded one-forms $dz^A$, $dc^i$ such that
\be
dz^A\w dc^i=-dc^i\w dz^A, \qquad dc^i\w dc^j= dc^j\w dc^i.
\ee
Accordingly, the graded Chevalley--Eilenberg coboundary operator
$d$ (\ref{ws86}), called the graded exterior differential, reads
\be
d\f= dz^A \w \dr_A\f +dc^a\w \frac{\dr}{\dr c^a}\f,
\ee
where derivatives $\dr_\la$, $\dr/\dr c^a$ act on coefficients of
graded exterior forms by the formula (\ref{cmp50a}), and they are
graded commutative with graded forms $dz^A$ and $dc^a$. The
formulas (\ref{ws45}) -- (\ref{+117}) hold.

One can show that the differential bigraded algebra $\cS^*[E;Z]$
(\ref{33f21}) is a minimal differential calculus over $\cA_E$,
i.e., it is generated by elements $df$, $f\in \cA_E$
\cite{book09,sard13}.

The bigraded de Rham complex (\ref{t100}) of the minimal graded
Chevalley--Eilenberg differential calculus $\cS^*[E;Z]$ reads
\mar{+137}\beq
0\to \mathbb R\to \cA_E \ar^d \cS^1[E;Z]\ar^d\cdots
\cS^k[E;Z]\ar^d\cdots. \label{+137}
\eeq
Its cohomology $H^*(\cA_E)$  is called the de Rham cohomology of a
simple graded manifold $(Z,\gA_E)$.

In particular, given the differential graded algebra $\cO^*(Z)$ of
exterior forms on $Z$, there exists a canonical monomorphism
\mar{uut}\beq
\cO^*(Z)\to \cS^*[E;Z] \label{uut}
\eeq
and a body epimorphism $\cS^*[E;Z]\to \cO^*(Z)$ which are cochain
morphisms of the de Rham complex (\ref{+137}) and that of
$\cO^*(Z)$. The de Rham cohomology of a simple graded manifold
$(Z,\gA_E)$ equals the de Rham cohomology of its body $Z$
\cite{book09}.

\section{Graded bundles}

A morphism of graded manifolds $(Z,\gA) \to (Z',\gA')$ is defined
as that of local-ringed spaces
\mar{su1}\beq
\phi:Z\to Z', \qquad \wh\Phi: \gA'\to \phi_*\gA, \label{su1}
\eeq
where $\phi$ is a manifold morphism and $\wh\Phi$ is a sheaf
morphism of $\gA'$ to the direct image $\phi_*\gA$ of $\gA$ onto
$Z'$ \cite{book05,ten}. The morphism (\ref{su1}) of graded
manifolds is said to be:

$\bullet$ a monomorphism if $\phi$ is an injection and $\wh\Phi$
is an epimorphism;

$\bullet$ an epimorphism if $\phi$ is a surjection and $\wh\Phi$
is a monomorphism.

\begin{definition} \label{su23} \mar{su23} An epimorphism of graded manifolds $(Z,\gA) \to
(Z',\gA')$ where $Z\to Z'$ is a fibre bundle is called the graded
bundle.
\end{definition}

Let $E\to Z$ and $E'\to Z$ be vector bundles and $\Phi: E\to E'$
their bundle morphism over a morphism $\phi: Z\to Z'$. Then every
section $s^*$ of the dual bundle $E'^*\to Z'$ defines the
pull-back section $\Phi^*s^*$ of the dual bundle $E^*\to Z$ by the
law
\be
v_z\rfloor \Phi^*s^*(z)=\Phi(v_z)\rfloor s^*(\vf(z)), \qquad
v_z\in E_z.
\ee
It follows that a bundle morphism $(\Phi,\phi)$ yields a morphism
of simple graded manifolds
\mar{w901}\beq
(Z,\gA_E) \to (Z',\gA_{E'}). \label{w901}
\eeq
This is a pair $(\phi,\wh\Phi=\phi_*\circ\Phi^*)$ of a morphism
$\phi$ of  body manifolds and the composition $\phi_*\circ\Phi^*$
of the pull-back
\be
\cA_{E'}\ni f\to \Phi^*f\in\cA_E
\ee
of graded functions and the direct image $\phi_*$ of a sheaf
$\gA_E$ onto $Z'$. Relative to local bases $(z^A,c^a)$ and
$(z'^A,c'^a)$ for $(Z,\gA_E)$ and $(Z',\gA_{E'})$, the morphism
(\ref{w901}) of simple graded manifolds reads
\be
z'=\phi(z), \qquad \wh\Phi(c'^a)=\Phi^a_b(z)c^b.
\ee
For instance, the graded manifold morphism (\ref{w901}) is a
monomorphism (resp. epimorphism) if $\Phi$ is a bundle injection
(resp. surjection).

In particular, let $(Y,\gA)$ be a graded manifold whose body $Z=Y$
is a fibre bundle $\pi:Y\to X$. Let us consider the trivial graded
manifold $(X,\gA_0=C^\infty_X)$ (\ref{trv}). Then we have a graded
bundle
\mar{su3}\beq
(Y,\gA) \to (X,C^\infty_X) \label{su3}
\eeq
in accordance with Definition \ref{su23}.

\begin{definition} \label{su22} \mar{su22}
We agree to call the graded bundle (\ref{su3}) over a trivial
graded manifold $(X, C^\infty_X)$ the graded bundle over a smooth
manifold. Let us denote it by $(X,Y,\gA)$.
\end{definition}

If $Y\to X$ is a vector bundle, this is a particular case of
graded fibre bundles in \cite{hern,mont} when their base is a
trivial graded manifold.

\begin{remark} \label{su20} \mar{su20}
Let $Y\to X$ be a fibre bundle. Then a  trivial graded manifold
$(Y,C^\infty_Y)$ together with a real ring monomorphism
$C^\infty(X)\to C^\infty(Y)$ is the graded bundle
$(X,Y,C^\infty_Y)$ (\ref{su3}).
\end{remark}

\begin{remark} \label{su21} \mar{su21} A graded manifold $(X,\gA)$ itself can
be treated as the graded bundle $(X,X, \gA)$ (\ref{su3})
associated to the identity smooth bundle $X\to X$.
\end{remark}

Given a graded bundle $(X,Y,\gA)$, a local basis for a graded
manifold $(Y,\gA)$ can be brought into a form $(x^\la, y^i, c^a)$
where $(x^\la, y^i)$ are bundle coordinates of $Y\to X$.

Let $(Y,\gA_F)$ be a simple graded manifold modelled over a vector
bundle $F\to Y$. This is a graded bundle $(X,Y,\gA_F)$ together
with a composite bundle
\mar{su5}\beq
F\to Y\to X.  \label{su5}
\eeq
Accordingly, there is a composite bundle
\be
\op\w_Y F^*\to Y\to X
\ee
so that the structure ring $\cA_F=\w F^*(Y)$ (\ref{33f1}) of
graded functions on a simple graded manifold $(Y,\gA_F)$ is a
graded commutative $C^\infty(X)$-ring. Let the composite bundle
(\ref{su5}) be provided with adapted bundle coordinates
$(x^\la,y^i,q^a)$ possessing transition functions
\be
x'^\la(x^\mu), \qquad y'^i(x^\m,y^j), \qquad
q'^a=\rho^a_b(x^\mu,y^j)q^b.
\ee
Then the corresponding local basis for a graded manifold is
$(x^\la,y^i,c^a)$ together with transition functions
\be
x'^\la(x^\mu), \qquad y'^i(x^\m,y^j), \qquad
c'^a=\rho^a_b(x^\mu,j^j)c^b.
\ee
We call it the local basis for a graded bundle $(X,Y,\gA_F)$.

Any global section $s$ of a fibre bundle $Y\to X$ yields the
pull-back vector bundle $s^*F\to X$ over $X$ which is a subbundle
of a fibre bundle $Y\to X$ \cite{sard09,book13}. Let
$(X,\gA_{s^*F})$ be a simple graded manifold modelled over the
pull-back bundle $s^F$. Its structure ring consists of global
sections of the pull-back bundle
\mar{su9}\beq
s^*\w F^*=\w s^*F^*\to X, \label{su9}
\eeq
which is a subbundle of a fibre bundle $\w F^*\to X$. With a
global section $s$ of a fibre bundle $Y\to X$, any global section
$f$ of an exterior bundle $\w F^*\to Y$ defines a global section
$f\circ s$ of the exterior bundle (\ref{su9}). The correspondence
$f\to f\circ s$ yields an even morphism of a graded commutative
$C^\infty(X)$-ring $\cA_F$ to a graded commutative ring
$\cA_{s^*F}$. This morphism is an epimorphism because any global
section of a subbundle $s^*\w F^*\subset \w F^*\to X$ is a local
section of $\w F^*\to Y$ over a closed subset $s(X)\subset Y$ and,
therefore, it can be extended to a global section $f$ of $\w
F^*\to Y$. Passing to the sheafs $\gA_F$ and $\gA_{s^*F}$ of germs
of graded functions, we come to the following.

\begin{theorem} \label{su10} \mar{su10}
Given a graded bundle $(X, Y,\gA_F)$, any global section $s:X\to
Y$ of a fibre bundle $Y\to X$ yields a monomorphism of a graded
manifold $(X,\gA_{s^*F})$ to a graded manifold $(Y,\gA_F)$.
\end{theorem}

\section{Graded jet manifolds}

As was mentioned above, Lagrangian theory of even fields
represented by sections of a fibre bundle $Y\to X$ is formulated
on jet manifolds $J^*Y$ of $Y\to X$. Therefore, one must define
jets of odd fields.

Given a graded manifold $(X,\gA)$ and its structure ring $\cA$,
one can define the jet module $J^1\cA$ of a $C^\infty(X)$-ring
$\cA$ \cite{book05,book12}. If $(X,\gA_E)$ is a simple graded
manifold modelled over a vector bundle $E\to X$, the jet module
$J^1\cA_E$ is a module of global sections of the jet bundle
$J^1(\w E^*)$. From the mathematical viewpoint, a problem is that
$J^1\cA_E$ fails to be a structure ring of some graded manifold.
From the physical viewpoint, elements of $J^1\cA_E$ are affine in
jets of odd fields.

By this reason, we have suggested a different construction of jets
of graded manifolds, though it is applied only to simple graded
manifolds \cite{book09,book13,sard13}.

Let $(X,\cA_E)$ be a simple graded manifold modelled over a vector
bundle $E\to X$. Let us consider a $k$-order jet manifold $J^kE$
of $E$ It is a vector bundle over $X$. Then let $(X,\cA_{J^kE})$
be a simple graded manifold modelled over $J^kE\to X$.

\begin{definition} \label{su25} \mar{su25}
We agree to call $(X,\cA_{J^kE})$ the graded $k$-order jet
manifold of a simple graded manifold $(X,\cA_E)$.
\end{definition}

Given a splitting domain $(U; x^\la,c^a)$ of a graded manifold
$(Z,\cA_E)$, we have a splitting domain
\be
(U; x^\la,c^a, c^a_\la,c^a_{\la_1\la_2}, \ldots
c^a_{\la_1\ldots\la_k}), \qquad
c'{}^a_{\la\la_1\ldots\la_r}=\rho^a_b(x)c^a_{\la\la_1\ldots\la_r}
+ \dr_\la\rho^a_b(x) c^a_{\la_1\ldots\la_r},
\ee
of a graded jet manifold $(X,\cA_{J^kE})$.

\begin{remark} \label{su31} \mar{su31}
Given a manifold $X$ and its coordinate chart $(U; x^\la)$, a
multi-index $\La$ of the length $|\La|=k$ throughout denotes a
collection of indices $(\la_1...\la_k)$ modulo permutations. By
$\la+\La$ is meant a multi-index $(\la\la_1\ldots\la_k)$.
Summation over a multi-index $\La$ means separate summation over
each its index $\la_i$. We use the compact notation
$\dr_\La=\dr_{\la_k}\circ\cdots\circ\dr_{\la_1}$ and
$\La=(\la_1...\la_k)$.
\end{remark}

As was mentioned above, a graded manifold is a particular graded
bundle over its body (Remark \ref{su21}). Then Definition
\ref{su25} of graded jet manifolds is generalized to graded
bundles over smooth manifolds as follows.

Let $(X,Y,\gA_F)$ be a graded bundle modelled over the composite
bundle (\ref{su5}). It is readily observed that the jet manifold
$J^rF$ of $F\to X$ is a vector bundle $J^rF\to J^rY$ coordinated
by $(x^\la, y^i_\La, q^a_\La)$, $0\leq |\La|\leq r$. Let
$(J^rY,\gA_r=\gA_{J^rF})$ be a simple graded manifold modelled
over this vector bundle. Its local basis is $(x^\la, y^i_\La,
c^a_\La)$, $0\leq|\La|\leq r$.

\begin{definition} \label{su32} \mar{su32}
We call $(J^rY,\gA_r)$ the graded $r$-order jet manifold of a
graded bundle $(X,Y,\gA_F)$.
\end{definition}

In particular, let $Y\to X$ be a smooth bundle seen as a trivial
graded bundle $(X, Y, C^\infty_Y)$ modelled over a composite
bundle $Y\times{0}\to Y\to X$. Then its graded jet manifold in
accordance with Definition \ref{su32} is a trivial graded bundle
$(X, J^rY, C^\infty_{J^rY)})$, i.e., a jet manifold $J^rY$ of $Y$.

Thus, Definition \ref{su32} of graded jet manifolds of graded
bundles is compatible with the conventional definition of jets of
fibre bundles. As was mentioned above, it however differs from
that of jet graded bundles in \cite{hern,mont}.

Jet manifolds $J^*Y$ of a fibre bundle $Y\to X$ form the inverse
sequence (\ref{j1}) where $\pi^r_{r-1}$ are affine bundles. One
can think of elements of its projective limit $J^\infty Y$ as
being infinite order jets of sections of $Y\to X$ identified by
their Taylor series at points of $X$. The set $J^\infty Y$ is
endowed with the projective limit topology which makes $J^\infty
Y$ into a paracompact Fr\'echet manifold \cite{book09,tak2}. It is
called the infinite order jet manifold and is provided with with
the adapted coordinates
\mar{j3}\beq
 (x^\la, y^i_\La), \quad 0\leq|\La|, \qquad
{y'}^i_{\la+\La}=\frac{\dr x^\m}{\dr x'^\la}d_\m y'^i_\La,
\label{j3}
\eeq
where
\be
d_\la= \dr_\la + y^i_\la \dr_i +
\op\sum_{0<|\La|}y^i_{\la+\La}\dr^\La_i
\ee
are total derivatives. The inverse sequence (\ref{j1}) of jet
manifolds yields the direct sequence of graded differential
algebras $\cO_r^*$ of exterior forms on finite order jet manifolds
\mar{5.7}\beq
\cO^*(X)\op\longrightarrow^{\pi^*} \cO^*(Y)
\op\longrightarrow^{\pi^1_0{}^*} \cO_1^* \longrightarrow \cdots
\cO_{r-1}^*\op\longrightarrow^{\pi^r_{r-1}{}^*}
 \cO_r^* \longrightarrow\cdots, \label{5.7}
\eeq
where $\pi^r_{r-1}{}^*$ are the pull-back monomorphisms. Its
direct limit $\cO^*_\infty =\op\lim^\to \cO_r^*$ exists and
consists of all exterior forms on finite order jet manifolds
modulo the pull-back identification. The $\cO^*_\infty$ is a
differential graded algebra which inherits the operations of the
exterior differential $d$ and exterior product $\w$ of exterior
algebras $\cO^*_r$. A key point is that the cohomology
$H^*(\cO_\infty^*)$ of the de Rham complex
\mar{5.13} \beq
0\longrightarrow \mathbb R\longrightarrow \cO^0_\infty
\op\longrightarrow^d\cO^1_\infty \op\longrightarrow^d \cdots
\label{5.13}
\eeq
of a differential graded algebra $\cO^*_\infty$ equals the de Rham
cohomology of a fibre bundle $Y$ \cite{and,book09}.

Surjections $\pi^{r+1}_r:J^{r+1}Y\to J^rY$ (\ref{j1}) yield
epimorphisms of graded manifolds
\be
(\pi^{r+1}_r,\wh \pi^{r+1}_r):(J^{r+1}Y,\gA_{r+1}) \to
(J^rY,\gA_r),
\ee
including the sheaf monomorphisms
\mar{su50}\beq
\wh \pi^{r+1}_r:\pi_r^{r+1*}\gA_r\to \gA_{r+1}, \label{su50}
\eeq
where $\pi_r^{r+1*}\gA_r$ is the pull-back onto $J^{r+1}Y$ of a
continuous fibre bundle $\gA_r\to J^rY$. As a consequence, we have
the inverse sequence of graded manifolds (\ref{su14}).

The sheaf monomorphism (\ref{su50}) induces a monomorphism of
canonical presheaves $\ol \gA_r\to \ol \gA_{r+1}$, which
associates to each open subset $U\subset J^{r+1}Y$ the ring of
sections of $\gA_r$ over $\pi^{r+1}_r(U)$. Accordingly, there is a
monomorphism of the structure rings
\mar{34f1}\beq
\pi_r^{r+1*}:\cS^0_r[F;Y]\to \cS^0_{r+1}[F;Y] \label{34f1}
\eeq
of graded functions on graded manifolds $(J^rY,\gA_r)$ and
$(J^{r+1}Y,\gA_{r+1})$. Since the differential calculus
$\cS^*_r[F;Y]$ and $\cS^*_{r+1}[F;Y]$ are minimal. Therefore, the
monomorphism (\ref{34f1}) yields a monomorphism of differential
bigraded algebras
\mar{v4'}\beq
\pi_r^{r+1*}:\cS^*_r[F;Y]\to \cS^*_{r+1}[F;Y]. \label{v4'}
\eeq
As a consequence, we have the direct system of differential
bigraded algebras
\mar{j2}\beq
\cS^*[F;Y]\ar^{\pi^*} \cS^*_1[F;Y]\ar\cdots \cS^*_{r-1}[F;Y]
\op\ar^{\pi^{r*}_{r-1}}\cS^*_r[F;Y]\ar\cdots. \label{j2}
\eeq
The differential bigraded algebra  $\cS^*_\infty[F;Y]$ that we
associate to a graded bundle $(Y,\gA_F)$ is defined as the direct
limit
\mar{5.77a}\beq
\cS^*_\infty [F;Y]=\op\lim^\to \cS^*_r[F;Y] \label{5.77a}
\eeq
of the direct system (\ref{j2}). It consists of all graded
exterior forms $\f\in \cS^*[F_r;J^rY]$ on graded manifolds
$(J^rY,\gA_r)$ modulo the monomorphisms (\ref{v4'}). Its elements
obey the relations (\ref{ws45}).

The cochain monomorphisms $\cO^*_r\to \cS^*_r[F;Y]$ (\ref{uut})
provide a monomorphism of the direct system (\ref{5.7}) to the
direct system (\ref{j2}) and, consequently, a monomorphism
$\cO^*_\infty\to \cS^*_\infty[F;Y]$ of their direct limits. In
particular, $\cS^*_\infty[F;Y]$ is an $\cO^0_\infty$-algebra.

One can think of  elements of $\cS^*_\infty[F;Y]$ as being graded
differential forms on an infinite order jet manifold $J^\infty Y$
\cite{book09}. In particular, one can restrict $\cS^*_\infty[F;Y]$
to the coordinate chart (\ref{j3}) of $J^\infty Y$ and say that
$\cS^*_\infty[F;Y]$ as an $\cO^0_\infty$-algebra is locally
generated by  the elements
\be
(c^a_\La,
dx^\la,\thh^a_\La=dc^a_\La-c^a_{\la+\La}dx^\la,\thh^i_\La=
dy^i_\La-y^i_{\la+\La}dx^\la), \qquad 0\leq |\La|,
\ee
where $c^a_\La$, $\thh^a_\La$ are odd and $dx^\la$, $\thh^i_\La$
are even. We agree to call $(y^i,c^a)$ the local generating basis
for $\cS^*_\infty[F;Y]$. Let the collective symbol $s^A$ stand for
its elements. Accordingly, the notation $s^A_\La$ and
$\thh^A_\La=ds^A_\La- s^A_{\la+\La}dx^\la$ is introduced. For the
sake of simplicity, we further denote $[A]=[s^A]$.

\section{Lagrangian theory of even and odd fields}

A differential bigraded algebra $\cS^*_\infty[F;Y]$ is decomposed
into $\cS^0_\infty[F;Y]$-modules $\cS^{k,r}_\infty[F;Y]$ of
$k$-contact and $r$-horizontal graded forms together with the
corresponding projections
\be
h_k:\cS^*_\infty[F;Y]\to \cS^{k,*}_\infty[F;Y], \qquad
h^m:\cS^*_\infty[F;Y]\to \cS^{*,m}_\infty[F;Y].
\ee
Accordingly, the graded exterior differential $d$ on
$\cS^*_\infty[F;Y]$ falls into a sum $d=d_V+d_H$ of the vertical
graded differential
\be
d_V \circ h^m=h^m\circ d\circ h^m, \qquad d_V(\f)=\thh^A_\La \w
\dr^\La_A\f, \qquad \f\in\cS^*_\infty[F;Y],
\ee
and the total graded differential
\be
d_H\circ h_k=h_k\circ d\circ h_k, \qquad d_H\circ h_0=h_0\circ d,
\qquad d_H(\f)=dx^\la\w d_\la(\f),
\ee
where
\be
d_\la = \dr_\la + \op\sum_{0\leq|\La|} s^A_{\la+\La}\dr_A^\La
\ee
are the graded total derivatives. These differentials obey the
nilpotent relations
\be
d_H\circ d_H=0, \qquad d_V\circ d_V=0, \qquad d_H\circ
d_V+d_V\circ d_H=0.
\ee

Similarly to a differential graded algebra $\cO^*_\infty$, a
differential bigraded algebra $\cS^*_\infty[F;Y]$ is provided with
the graded projection endomorphism
\be
&& \vr=\op\sum_{k>0} \frac1k\ol\vr\circ h_k\circ h^n:
\cS^{*>0,n}_\infty[F;Y]\to \cS^{*>0,n}_\infty[F;Y], \\
&& \ol\vr(\f)= \op\sum_{0\leq|\La|} (-1)^{\nm\La}\thh^A\w
[d_\La(\dr^\La_A\rfloor\f)], \qquad \f\in \cS^{>0,n}_\infty[F;Y],
\ee
such that $\vr\circ d_H=0$, and with the nilpotent graded
variational operator
\mar{34f10}\beq
\dl=\vr\circ d \cS^{*,n}_\infty[F;Y]\to
\cS^{*+1,n}_\infty[F;Y].\label{34f10}
\eeq
With these operators a differential bigraded algebra
$\cS^{*,}_\infty[F;Y]$ is decomposed into the Grassmann-graded
variational bicomplex \cite{book09,ijgmmp07,sard13}.

 We restrict our
consideration to its short variational subcomplex
\mar{g111}\ben
&& 0\to \mathbb R\to \cS^0_\infty[F;Y]\ar^{d_H}\cS^{0,1}_\infty[F;Y]
\cdots \ar^{d_H}\cS^{0,n}_\infty[F;Y]\ar^\dl \bE_1, \label{g111}\\
&& \bE_1=\vr(\cS^{1,n}_\infty[F;Y]), \qquad n=\di X, \nonumber
\een
and the subcomplex of one-contact graded forms
\mar{g112}\beq
 0\to \cS^{1,0}_\infty[F;Y]\ar^{d_H} \cS^{1,1}_\infty[F;Y]\cdots
\ar^{d_H}\cS^{1,n}_\infty[F;Y]\ar^\vr \bE_1\to 0. \label{g112}
\eeq

They possess the following cohomology
\cite{cmp04,ijgmmp07,sard13}.

\begin{theorem} \label{v11} \mar{v11}
Cohomology of the complex (\ref{g111}) equals the de Rham
cohomology $H^*_{DR}(Y)$ of $Y$. The complex (\ref{g112}) is
exact.
\end{theorem}

Decomposed into a variational bicomplex, the differential bigraded
algebra $\cS^*_\infty[F;Y]$ describes Grassmann-graded field
theory on a graded bundle $(X,Y,\gA_F)$. Its graded Lagrangian is
defined as an element
\be
L=\cL\om\in \cS^{0,n}_\infty[F;Y]
\ee
of the graded variational complex (\ref{g111}). Accordingly, a
graded exterior form
\mar{0709'}\beq
\dl L= \thh^A\w \cE_A\om=\op\sum_{0\leq|\La|}
 (-1)^{|\La|}\thh^A\w d_\La (\dr^\La_A L)\om\in \bE_1 \label{0709'}
\eeq
is said to be its graded Euler--Lagrange operator. We agree to
call a pair $(\cS^{0,n}_\infty[F;Y],L)$ the  Grassmann-graded (or,
simply, graded) Lagrangian system and $\cS^*_\infty[F;Y]$ the
field system algebra.

The following is a corollary of Theorem \ref{v11}
\cite{book09,sard13}.

\begin{theorem} \label{cmp26} \mar{cmp26}
Every $d_H$-closed graded form $\f\in\cS^{0,m<n}_\infty[F;Y]$
falls into the sum
\mar{g214}\beq
\f=h_0\si + d_H\xi, \qquad \xi\in \cS^{0,m-1}_\infty[F;Y],
\label{g214}
\eeq
where $\si$ is a closed $m$-form on $Y$. Any $\dl$-closed (i.e.,
variationally trivial) graded Lagrangian $L\in
\cS^{0,n}_\infty[F;Y]$ is a sum
\mar{g215}\beq
L=h_0\si + d_H\xi, \qquad \xi\in \cS^{0,n-1}_\infty[F;Y],
\label{g215}
\eeq
where $\si$ is a closed $n$-form on $Y$.
\end{theorem}

\begin{corollary} \mar{34c5} \label{34c5}
Any variationally trivial odd Lagrangian is $d_H$-exact.
\end{corollary}

The exactness of the complex (\ref{g112}) at the term
$\cS^{1,n}_\infty[F;Y]$ results in the following
\cite{cmp04,book09,sard13}.

\begin{theorem} \label{g103} \mar{g103}
Given a graded Lagrangian $L$, there is the decomposition
\mar{g99,'}\ben
&& dL=\dl L - d_H\Xi_L,
\qquad \Xi\in \cS^{n-1}_\infty[F;Y], \label{g99}\\
&& \Xi_L=L+\op\sum_{s=0} \thh^A_{\nu_s\ldots\nu_1}\w
F^{\la\nu_s\ldots\nu_1}_A\om_\la, \label{g99'}\\
&& F_A^{\nu_k\ldots\nu_1}= \dr_A^{\nu_k\ldots\nu_1}\cL-d_\la
F_A^{\la\nu_k\ldots\nu_1} +\si_A^{\nu_k\ldots\nu_1},\qquad
k=1,2,\ldots,\nonumber
\een
where local graded functions $\si$ obey the relations
\be
\si^\nu_A=0,\qquad \si_A^{(\nu_k\nu_{k-1})\ldots\nu_1}=0.
\ee
\end{theorem}

The form $\Xi_L$ (\ref{g99'}) provides a global Lepage equivalent
of a graded Lagrangian $L$. In particular, one can locally choose
$\Xi_L$ (\ref{g99'}) where all functions $\si$ vanish.

Given a graded Lagrangian system $(\cS^*_\infty[F;Y], L)$, by its
infinitesimal transformations are meant contact graded derivations
of the real graded commutative ring $\cS^0_\infty[F;Y]$. They
constitute a $\cS^0_\infty[F;Y]$-module $\gd \cS^0_\infty[F;Y]$
which is a real Lie superalgebra with respect to the Lie
superbracket (\ref{ws14}).

\begin{theorem} \label{35t1} \mar{35t1}
The derivation module $\gd\cS^0_\infty[F;Y]$ is isomorphic to the
$\cS^0_\infty[F;Y]$-dual $(\cS^1_\infty[F;Y])^*$ of the module of
graded one-forms $\cS^1_\infty[F;Y]$. It follows that the
differential bigraded algebra $\cS^*_\infty[F;Y]$ is minimal
differential calculus over the real graded commutative ring
$\cS^0_\infty[F;Y]$ \cite{book09,sard13}.
\end{theorem}

Let $\vt\rfloor\f$, $\vt\in \gd\cS^0_\infty[F;Y]$, $\f\in
\cS^1_\infty[F;Y]$, denote the corresponding interior product.
Extended to the differential bigraded algebra $\cS^*_\infty[F;Y]$,
it obeys the rule
\be
\vt\rfloor(\f\w\si)=(\vt\rfloor \f)\w\si
+(-1)^{|\f|+[\f][\vt]}\f\w(\vt\rfloor\si), \qquad \f,\si\in
\cS^*_\infty[F;Y].
\ee

Restricted to a coordinate chart (\ref{j3}) of $J^\infty Y$, the
algebra $\cS^*_\infty[F;Y]$ is a free $\cS^0_\infty[F;Y]$-module
generated by one-forms $dx^\la$, $\thh^A_\La$. Due to the
isomorphism stated in Theorem \ref{35t1}, any graded derivation
$\vt\in\gd\cS^0_\infty[F;Y]$ takes a local form
\mar{gg3}\beq
\vt=\vt^\la \dr_\la + \vt^A\dr_A + \op\sum_{0<|\La|}\vt^A_\La
\dr^\La_A, \label{gg3}
\eeq
where $\dr^\La_A\rfloor dy_\Si^B=\dl_A^B\dl^\La_\Si$ up to
permutations of multi-indices $\La$ and $\Si$.

Every graded derivation $\vt$ (\ref{gg3}) yields the graded Lie
derivative
\be
&& \bL_\vt\f=\vt\rfloor d\f+ d(\vt\rfloor\f), \qquad \f\in
\cS^*_\infty[F;Y],\\
&& \bL_\vt(\f\w\si)=\bL_\vt(\f)\w\si
+(-1)^{[\vt][\f]}\f\w\bL_\vt(\si),
\ee
of the differential bigraded algebra $\cS^*_\infty[F;Y]$. A graded
derivation $\vt$ (\ref{gg3}) is called contact if the Lie
derivative $\bL_\vt$ preserves the ideal of contact graded forms
of the differential bigraded algebra $\cS^*_\infty[F;Y]$.

With respect to the local generating basis $(s^A)$ for the
differential bigraded algebra $\cS^*_\infty[F;Y]$, any its contact
graded derivation takes a form
\mar{g105}\beq
\vt=\up_H+\up_V=\up^\la d_\la + [\up^A\dr_A +\op\sum_{|\La|>0}
d_\La(\up^A-s^A_\m\up^\m)\dr_A^\La], \label{g105}
\eeq
where $\up_H$ and $\up_V$ denotes the horizontal and vertical
parts of $\vt$. In particular, a vertical contact graded
derivation reads
\mar{j40}\beq
\vt= \up^A\dr_A +\op\sum_{|\La|>0} d_\La\up^A\dr_A^\La.
\label{j40}
\eeq

A glance at the expression (\ref{g105}) shows that a contact
graded derivation $\vt$ as an infinite order jet prolongation of
its restriction
\mar{jj15}\beq
\up=\up^\la\dr_\la +\up^A\dr_A \label{jj15}
\eeq
to the graded commutative ring $S^0[F;Y]$. We call $\up$
(\ref{jj15}) the generalized graded vector field.

A corollary of the decomposition (\ref{g99}) is the first
variational formula for a graded Lagrangian
\cite{jmp05,cmp04,book09}.

\begin{theorem} \label{j44} \mar{j44}
The Lie derivative of a graded Lagrangian along any contact graded
derivation (\ref{g105}) obeys the first variational formula
\mar{g107}\beq
\bL_\vt L= \up_V\rfloor\dl L +d_H(h_0(\vt\rfloor \Xi_L)) + d_V
(\up_H\rfloor\om)\cL, \label{g107}
\eeq
where $\Xi_L$ is the Lepage equivalent (\ref{g99'}) of $L$.
\end{theorem}

A contact graded derivation $\vt$ (\ref{g105}) is called a
variational symmetry of a graded Lagrangian $L$ if the Lie
derivative $\bL_\vt L$ is $d_H$-exact, i.e., $\bL_\vt L=d_H\si$.
Then a corollary of the first variational formula (\ref{g107}) is
the first Noether theorem for graded Lagrangians.

\begin{theorem} \label{j45} \mar{j45} If a contact graded derivation $\vt$
(\ref{g105}) is a variational symmetry of a graded Lagrangian $L$,
the first variational formula (\ref{g107}) leads to the weak
conservation law
\mar{35f2}\beq
0\ap d_H(h_0(\vt\rfloor\Xi_L)-\si) \label{35f2}
\eeq
on the shell Ker$\,\dl L$.
\end{theorem}

For the sake of brevity, the common symbol $\up$ further stands
for the generalized graded vector field $\up$ (\ref{jj15}), the
contact graded derivation $\vt$ (\ref{g105}) determined by $\up$,
and a Lie derivative $\bL_\vt$.

A vertical contact graded derivation $\up= \up^A\dr_A$ is said to
be nilpotent if $\up(\up\f)=0$ for any horizontal graded form
$\f\in S^{0,*}_\infty[F,Y]$. It is nilpotent only if it is odd and
iff the equality $\up(\up^A)=0$ holds for all $\up^A$
\cite{cmp04}.

\begin{remark} \label{rr35}
For the sake of convenience, right derivations $\op\up^\lto
=\rdr_A\up^A$ also are  considered. They act on graded functions
and differential forms $\f$ on the right by the rules
\be
\op\up^\lto(\f)=d\f\lfloor\op\up^\lto +d(\f\lfloor\op\up^\lto),
\qquad \op\up^\lto(\f\w\f')=(-1)^{[\f']}\op\up^\lto(\f)\w\f'+
\f\w\op\up^\lto(\f').
\ee
\end{remark}

\section{Lagrangian BRST theory}

As was mentioned above, quantization of Lagrangian field theory
essentially depends on its degeneracy characterized by a hierarchy
of Noether identities, and a first step to this quantization is
BRST extension of an original Lagrangian theory
\cite{barn,book09,gom}.

Any Lagrangian system admits Noether identities. Therefore, a
problem lies in separation of trivial and non-trivial Noether
identities that one can perform in the homology terms so that
trivial Noether identities are boundaries, whereas the non-trivial
ones are cycles of some chain complex \cite{jmp05a,lmp08,book09}.

Without a lose of generality, let a Lagrangian $L$ of a graded
Lagrangian system $(\cS^{0,n}_\infty[F;Y],L)$ be even. Its
Euler--Lagrange Euler--Lagrange operator $\dl L\in \bE_1\subset
\cS^{1,n}_\infty[F;Y]$ (\ref{0709'}) takes the values into a
vector bundle
\mar{41f33}\beq
\ol{VF}=V^*F\op\ot_F\op\w^n T^*X\to F, \label{41f33}
\eeq
where $V^*F$ is the vertical cotangent bundle of $F\to X$. For our
purpose of studying gauge theory, it suffices to consider a
pull-back composite bundle $F$ (\ref{su5}) that is
\mar{41f1}\beq
F=Y\op\times_X F^1\to Y\to X, \label{41f1}
\eeq
where $F^1\to X$ is a vector bundle.

Let us introduce the following notation. Given the vertical
tangent bundle $VE$ of a fibre bundle $E\to X$, by its
density-dual bundle is meant a fibre bundle
\mar{41f2}\beq
\ol{VE}=V^*E\op\ot_E \op\w^n T^*X. \label{41f2}
\eeq
If $E\to X$ is a vector bundle, we have
\be
\ol{VE}=\ol E\op\times_X E, \qquad \ol E=E^*\op\ot_X\op\w^n T^*X,
\ee
where $\ol E$ is called the density-dual of $E$. Let
$E=E^0\oplus_X E^1$ be a graded vector bundle over $X$. Its graded
density-dual is defined to be $\ol E=\ol E^1\oplus_X \ol E^0$. In
these terms, we treat the composite bundle $F$ (\ref{su5}) as a
graded vector bundle over $Y$ possessing only an odd part. The
density-dual $\ol{VF}$ (\ref{41f2}) of the vertical tangent bundle
$VF$ of $F\to X$ is $\ol{VF}$ (\ref{41f33}). If $F$ (\ref{su5}) is
the pull-back bundle (\ref{41f1}), then
\mar{41f4}\beq
\ol{VF}=((\ol F^1\op\oplus_Y V^*Y)\op\ot_Y\op\w^n T^*X)\op\oplus_Y
F^1 \label{41f4}
\eeq
is a graded vector bundle over $Y$. Given a graded vector bundle
$E=E^0\oplus_Y E^1$ over $Y$, we consider a composite bundle $E\to
E^0\to X$ and the differential  bigraded algebra (\ref{5.77a}):
\mar{41f5}\beq
\cP^*_\infty[E;Y]=\cS^*_\infty[E;E^0]. \label{41f5}
\eeq

Turn now to Noether identities of a Lagrangian system. One can
associate to any graded Lagrangian system $(\cS^*_\infty[F;Y],L)$
the chain complex (\ref{v042}) whose one-boundaries vanish
on-shell as follows.

Let us consider the density-dual $\ol{VF}$ (\ref{41f4}) of the
vertical tangent bundle $VF\to F$, and let us enlarge the original
field system algebra $\cS^*_\infty[F;Y]$ to the differential
bigraded algebra $\cP^*_\infty[\ol{VF};Y]$ (\ref{41f5}) with a
local generating basis
\be
(s^A, \ol s_A), \qquad [\ol s_A]=([A]+1){\rm mod}\,2.
\ee
Following the conventional terminology of Lagrangian BRST theory
\cite{barn,gom}, we agree to call its elements $\ol s_A$ the
antifields of antifield number Ant$[\ol s_A]= 1$. The differential
bigraded algebra $\cP^*_\infty[\ol{VF};Y]$ is endowed with the
nilpotent right graded derivation
\be
\ol\dl=\rdr^A \cE_A,
\ee
where $\cE_A$ are the variational derivatives (\ref{0709'}). Then
we have the chain complex
\mar{v042}\beq
0\lto \im\ol\dl \llr^{\ol\dl} \cP^{0,n}_\infty[\ol{VF};Y]_1
\llr^{\ol\dl} \cP^{0,n}_\infty[\ol{VF};Y]_2 \label{v042}
\eeq
of graded densities of antifield number $\leq 2$. Its
one-boundaries $\ol\dl\Phi$, $\Phi\in
\cP^{0,n}_\infty[\ol{VF};Y]_2$, by very definition, vanish
on-shell.

Any one-cycle
\mar{0712}\beq
\Phi= \op\sum_{0\leq|\La|} \Phi^{A,\La}\ol s_{\La A} \om \in
\cP^{0,n}_\infty[\ol{VF};Y]_1\label{0712}
\eeq
of the complex (\ref{v042}) is a differential operator on the
bundle $\ol{VF}$ such that it is linear on fibres of $\ol{VF}\to
F$ and its kernel contains the graded Euler--Lagrange operator
$\dl L$ (\ref{0709'}), i.e.,
\mar{0713}\beq
\ol\dl\Phi=0, \qquad \op\sum_{0\leq|\La|} \Phi^{A,\La}d_\La
\cE_A\om=0. \label{0713}
\eeq
Thus the one-cycles (\ref{0712}) define the Noether identities
(\ref{0713}) of the Euler--Lagrange operator $\dl L$ a graded
Lagrangian system $(\cS^*_\infty[F;Y],L)$.

In particular, one-chains $\Phi$ (\ref{0712}) are necessarily
Noether identities if they are boundaries. Therefore, these
Noether identities are called trivial. Accordingly, non-trivial
Noether identities modulo the trivial ones are associated to
elements of the first homology $H_1(\ol\dl)$ of the complex
(\ref{v042}). A Lagrangian system is called degenerate if there
are non-trivial Noether identities.

Let us assume that a $C^\infty(X)$-module $H_1(\ol \dl)$ is
finitely generated. Namely, there exists a graded projective
$C^\infty(X)$-module $\cC_{(0)}\subset H_1(\ol \dl)$ of finite
rank possessing a local basis $\{\Delta_r\om\}$:
\mar{41f7}\beq
\Delta_r\om=\op\sum_{0\leq|\La|} \Delta_r^{A,\La}\ol s_{\La
A}\om,\qquad \Delta_r^{A,\La}\in \cS^0_\infty[F;Y], \label{41f7}
\eeq
such that any element $\Phi\in H_1(\ol \dl)$ factorizes as
\mar{xx2}\beq
\Phi= \op\sum_{0\leq|\Xi|} \Phi^{r,\Xi} d_\Xi \Delta_r \om, \qquad
\Phi^{r,\Xi}\in \cS^0_\infty[F;Y], \label{xx2}
\eeq
through elements (\ref{41f7}) of $\cC_{(0)}$. Thus, all
non-trivial Noether identities (\ref{0713}) result from the
Noether identities
\mar{v64}\beq
\ol\dl\Delta_r= \op\sum_{0\leq|\La|} \Delta_r^{A,\La} d_\La
\cE_A=0, \label{v64}
\eeq
called the complete Noether identities.

Hereafter, we restrict our consideration to Lagrangian systems
whose non-trivial Noether identities are finitely generated
because this is just the case of gauge theory (Section 9).

\begin{lemma}
If the homology $H_1(\ol\dl)$ of the complex (\ref{v042}) is
finitely generated in the above mentioned sense, this complex can
be extended to the one-exact chain complex (\ref{v66}) with a
boundary operator whose nilpotency conditions are equivalent to
the complete Noether identities (\ref{v64}).
\end{lemma}

Indeed, by virtue of Serre--Swan Theorem \ref{vv0}, the graded
module $\cC_{(0)}$ is isomorphic to a module of sections of the
density-dual $\ol E_0$ of some graded vector bundle $E_0\to X$.
Let us enlarge $\cP^*_\infty[\ol{VF};Y]$ to the differential
bigraded algebra
\mar{41f14}\beq
\ol\cP^*_\infty\{0\}=\cP^*_\infty[\ol{VF}\op\oplus_Y \ol E_0;Y]
\label{41f14}
\eeq
possessing the local generating basis $(s^A,\ol s_A, \ol c_r)$
where $\ol c_r$ are  Noether antifields of Grassmann parity $[\ol
c_r]=([\Delta_r]+1){\rm mod}\,2$ and antifield number ${\rm
Ant}[\ol c_r]=2$. The differential bigraded algebra (\ref{41f14})
is provided with an odd right graded derivation
\mar{41f10}\beq
\dl_0=\ol\dl + \op\sum_{0\leq|\La|}\rdr^r\Delta_r^{A,\La}\ol
s_{\La A},  \label{41f10}
\eeq
called the Koszul--Tate operator. It is nilpotent iff the complete
Noether identities (\ref{v64}) hold.

Then $\dl_0$ (\ref{41f10}) is a boundary operator of a short chain
complex, called the Koszul--Tate complex,
\mar{v66}\beq
0\lto \im\ol\dl \op\lto^{\ol\dl}
\cP^{0,n}_\infty[\ol{VF};Y]_1\op\lto^{\dl_0}
\ol\cP^{0,n}_\infty\{0\}_2 \op\lto^{\dl_0}
\ol\cP^{0,n}_\infty\{0\}_3 \label{v66}
\eeq
of graded densities of antifield number $\leq 3$. Let $H_*(\dl_0)$
denote its homology. It is readily observed that
$H_0(\dl_0)=H_1(\dl_1)=0$. Let us consider the second homology
$H_2(\dl_0)$ of the complex (\ref{v66}). Its two-chains  read
\mar{41f9}\beq
\Phi= G + H= \op\sum_{0\leq|\La|} G^{r,\La}\ol c_{\La r}\om +
\op\sum_{0\leq|\La|,|\Si|} H^{(A,\La)(B,\Si)}\ol s_{\La A}\ol
s_{\Si B}\om. \label{41f9}
\eeq
Its two-cycles define the first-stage Noether identities
\be
\dl_0 \Phi=0, \qquad   \op\sum_{0\leq|\La|} G^{r,\La}d_\La\Delta_r
\om =-\ol\dl H.
\ee
They are trivial either if a two-cycle $\Phi$ (\ref{41f9}) is a
$\dl_0$-boundary or its summand $G$ vanishes on-shell. A field
theory is said to be irreducible if first-stage Noether identities
are trivial. As was mentioned above, we restrict our consideration
to this case.

The inverse second Noether theorem (Theorem \ref{w35}), that we
formulate in homology terms, associates to the Koszul--Tate
complex (\ref{v66}) of non-trivial Noether identities the cochain
sequence (\ref{w108}) with the ascent operator $u$ (\ref{w33})
whose components are non-trivial gauge symmetries of Lagrangian
system.

Let us start with the following notation. Given the differential
bigraded algebra $\ol\cP^*_\infty\{0\}$ (\ref{41f14}), we consider
a differential bigraded algebra
\mar{w5}\beq
\cP^*_\infty\{0\}=\cP^*_\infty[F\op\oplus_Y E_0;Y], \label{w5}
\eeq
possessing a local generating basis $(s^A, c^r)$, $[c^r]=([\ol
c_r]+1){\rm mod}\,2$, and a differential bigraded algebra
\mar{w6}\beq
P^*_\infty\{0\}=\cP^*_\infty[\ol{VF}\op\oplus_Y E_0\op\oplus_Y \ol
E_0;Y] \label{w6}
\eeq
with a local generating basis $(s^A, \ol s_A, c^r,\ol c_r)$. Their
elements $c^r$ are called the ghosts of ghost number gh$[c^r]=1$
and antifield number ${\rm Ant}[c^r]=-1$. A $C^\infty(X)$-module
$\cC^{(0)}$ of ghosts is the density-dual of a module $\cC_{(0)}$
of antifields. The differential bigraded algebras
$\ol\cP^*_\infty\{0\}$ (\ref{41f14}) and $\cP^*_\infty\{0\}$
(\ref{w5}) are subalgebras of $P^*_\infty\{0\}$ (\ref{w6}). The
operator $\dl_0$ (\ref{41f10}) is naturally extended to a graded
derivation of the differential bigraded algebra $P^*_\infty\{0\}$.

\begin{theorem} \label{w35} \mar{w35} Given the Koszul--Tate complex (\ref{v66}),
the module of graded densities $\cP_\infty^{0,n}\{0\}$ contains a
short cochain sequence
\mar{w108,w33}\ben
&& 0\to \cS^{0,n}_\infty[F;Y]\ar^u
\cP^{0,n}_\infty\{N\}^1, \label{w108}\\
&& u= u^A\frac{\dr}{\dr s^A}, \qquad u^A =\op\sum_{0\leq|\La|}
c^r_\La\eta(\Delta^A_r)^\La, \label{w33}
\een
of ghost number $\leq 1$. Its ascent operator $u$ (\ref{w33}) is
an odd graded derivation of ghost number 1 is a variational
symmetry of a graded Lagrangian $L$.
\end{theorem}

A glance at the expression (\ref{w33}) shows that the variational
symmetry $u$ is a linear differential operator on the
$C^\infty(X)$-module $\cC^{(0)}$ of ghosts with values into the
real space of variational symmetries. Therefore, $u$ (\ref{w33})
is a gauge symmetry of a graded Lagrangian $L$ which is associated
to the complete Noether identities (\ref{v64})
\cite{book09,gauge09}. It is called the gauge operator.

Being a variational symmetry, the gauge operator $u$ (\ref{w33})
defines the weak conservation law (\ref{35f2}). Let $u$ be an
exact Lagrangian symmetry. In this case, the associated symmetry
current
\mar{42f50}\beq
\cJ_u= -h_0(u\rfloor\Xi_L) \label{42f50}
\eeq
is conserved. The peculiarity of gauge conservation laws is that
the symmetry current (\ref{42f50}) is reduced to a superpotential
as follows \cite{book09,gauge09}.

\begin{theorem} \label{supp'} \mar{supp'}
If $u$ (\ref{w33}) is an exact gauge symmetry of a graded
Lagrangian $L$, the corresponding conserved symmetry current
$\cJ_u$ (\ref{42f50}) takes a form
\mar{b381'}\beq
\cJ_u= W+ d_HU=(W^\m +d_\nu U^{\nu\m})\om_\m, \label{b381'}
\eeq
where the term $W$ vanishes on-shell, i.e., $W\ap 0$, and $U$ is a
horizontal graded $(n-2)$-form.
\end{theorem}

In contrast with the Koszul--Tate operator $\dl_0$ (\ref{41f10}),
the gauge operator $u$ (\ref{w33}) need not be nilpotent. Let us
study its extension to a nilpotent graded derivation
\mar{w109}\beq
\bbc=u+ \g= u +\g^r\frac{\dr}{\dr c^r} \label{w109}
\eeq
of ghost number 1 by means of antifield-free terms $\g$ of higher
polynomial degree in ghosts $c^r$ and their jets $c^r_\La$. We
call $\bbc$ (\ref{w109}) the BRST operator, where gauge symmetries
are extended to BRST transformations acting on ghosts
\cite{jmp09,book09}.

We refer to \cite{book09} for necessary conditions of the
existence of such a BRST extension. If such an extension exists, a
Lagrangian $L$ also admits a certain BRST extension as follows.

The differential bigraded algebra $P^*_\infty\{0\}$ (\ref{w6}) is
a particular field-antifield theory of the following type
\cite{barn,lmp08,gom}. Let us consider a pull-back composite
bundle
\be
W=Z\op\times_X Z'\to Z\to X
\ee
where $Z'\to X$ is a vector bundle. Let us regard it as a graded
vector bundle over $Z$ possessing only odd part. The density-dual
$\ol{VW}$ of the vertical tangent bundle $VW$ of $W\to X$ is a
graded vector bundle
\be
\ol{VW}=((\ol Z'\op\oplus_Z V^*Z)\op\ot_Z\op\w^n T^*X)\op\oplus_Y
Z'
\ee
over $Z$ (cf. (\ref{41f4})). Let us consider the differential
bigraded algebra $\cP^*_\infty[\ol{VW};Z]$ (\ref{41f5}) with the
local generating basis
\be
(z^a,\ol z_a), \qquad [\ol z_a]=([z^a]+1){\rm mod}\,2.
\ee
Its elements $z^a$ and $\ol z_a$ are called fields and antifields,
respectively.

Graded densities of this differential bigraded algebra are endowed
with the antibracket
\mar{f11}\beq
\{\gL \om,\gL'\om\}=\left[\frac{\op\dl^\lto \gL}{\dl \ol
z_a}\frac{\dl \gL'}{\dl z^a} +
(-1)^{[\gL']([\gL']+1)}\frac{\op\dl^\lto \gL'}{\dl \ol
z_a}\frac{\dl \gL}{\dl z^a}\right]\om. \label{f11}
\eeq
With this antibracket, one associates to any even Lagrangian
$\gL\om$ the odd vertical graded derivations
\mar{w37,lmp1}\ben
&&\up_\gL=\op\cE^\lto{}^a\dr_a=\frac{\op\dl^\lto \gL}{\dl \ol z_a}
\frac{\dr}{\dr z^a}, \label{w37}\\
&&\ol\up_\gL=\rdr^a\cE_a=\frac{\op\dr^\lto}{\dr \ol z_a}\frac{\dl
\gL}{\dl z^a}, \label{w37'}\\
&& \vt_\gL=\up_\gL+ \ol\up_\gL^l=(-1)^{[a]+1}\left(\frac{\dl \gL}{\dl
\ol z^a}\frac{\dr}{\dr z_a}+\frac{\dl \gL}{\dl z^a}\frac{\dr}{\dr
\ol z_a}\right),  \label{lmp1}
\een
such that $\vt_\gL(\gL'\om)=\{\gL\om,\gL' \om\}$.

\begin{theorem} \label{w39} \mar{w39} The following conditions are
equivalent \cite{book09}.

(i) The antibracket of a Lagrangian $\gL\om$ is $d_H$-exact, i.e.,
\mar{w44}\beq
\{\gL\om,\gL\om\}=2\frac{\op\dl^\lto \gL}{\dl \ol z_a}\frac{\dl
\gL}{\dl z^a}\om =d_H\si. \label{w44}
\eeq

(ii) The graded derivations $\up$ (\ref{w37}) and $\ol\up$
(\ref{w37'}) are variational symmetries of a Lagrangian $\gL\om$.

(iii) The graded derivation $\vt_\gL$ (\ref{lmp1}) is nilpotent.
\end{theorem}

The equality (\ref{w44}) is called the classical master equation.
A solution of the master equation (\ref{w44}) is called
non-trivial if both the derivations (\ref{w37}) and (\ref{w37'})
do not vanish. Being an element of the differential bigraded
algebra $P^*_\infty\{0\}$ (\ref{w6}), an original Lagrangian $L$
obeys the master equation (\ref{w44}) and yields the graded
derivations $\up_L=0$ (\ref{w37}) and $\ol\up_L=\ol\dl$
(\ref{w37'}), i.e., it is a trivial solution of the master
equation. One can show the following \cite{book09}.

\begin{theorem} \label{w130} \mar{w130} The master equation has a non-trivial
solution
\mar{w133}\beq
L_E=L+\bbc(c^r\ol c_r) \label{w133}
\eeq
if the gauge operator $u$ (\ref{w33}) can be extended to the BRST
operator $\bbc$ (\ref{w109}),
\end{theorem}

The non-trivial solution $L_E$ (\ref{w133}) of the master equation
is called the BRST extension of an original Lagrangian $L$. As was
mentioned above, it is a necessary step towards quantization of
classical Lagrangian field theory in terms of functional
integrals.

\section{Yang--Mills gauge theory}

Yang--Mills gauge theory is a theory of principal connections on
principal bundles \cite{book09,sard13}.

Let
\mar{51f1}\beq
\pi_P :P\to X \label{51f1}
\eeq
be a smooth principal bundle with a structure Lie group $G$ which
acts freely and transitively on $P$ on the right by a fibrewise
morphism
\mar{1}\beq
 R_{GP}: G\op\times_X P\ni  p\to pg\in P,  \qquad \pi_P(p)=\pi_P(pg), \qquad p\in P. \label{1}
\eeq
As a consequence, the quotient of $P$ with respect to the action
(\ref{1}) of $G$ is diffeomorphic to a base $X$, i.e., $P/G=X$.

A principal $G$-bundle $P$ is equipped with a bundle atlas
\mar{51f2}\beq
\Psi_P=\{(U_\al,\psi^P_\al),\vr_{\al\bt}\} \label{51f2}
\eeq
whose trivialization morphisms
\be
\psi_\al^P: \pi_P^{-1}(U_\al)\to U_\al\times G
\ee
obey the condition $\psi_\al^P(pg)=g\psi_\al^P(p)$, $g\in G$. Due
to this property, every trivialization morphism $\psi^P_\al$
determines a unique local section $z_\al:U_\al\to P$ such that
$(\psi^P_\al\circ z_\al)(x)=\bb$, $x\in U_\al$. The transformation
law for $z_\al$ reads
\mar{b1.202}\beq
z_\bt(x)=z_\al(x)\vr_{\al\bt}(x),\qquad x\in U_\al\cap
U_\bt.\label{b1.202}
\eeq
Conversely, the family $\{(U_\al,z_\al),\vr_{\al\bt}\}$ of local
sections of $P$ which obey the transformation law (\ref{b1.202})
uniquely determines a bundle atlas $\Psi_P$ of a principal bundle
$P$.

In particular, it follows that a principal bundle admits a global
section iff it is trivial.

\begin{example} \mar{52e1} \label{52e1} Let $H$ be a closed
subgroup of a real Lie group $G$. Then $H$ is a Lie group. Let
$G/H$ be the quotient of $G$ with respect to an action of $H$ on
$G$ by right multiplications. Then
\mar{ggh}\beq
\pi_{GH}:G\to G/H \label{ggh}
\eeq
is a principal $H$-bundle. If $H$ is a maximal compact subgroup of
$G$, then $G/H$ is diffeomorphic to $\mathbb R^m$ and, by virtue
of the well known theorem, $G\to G/H$ is a trivial bundle, i.e.,
$G$ is diffeomorphic to the product $\mathbb R^m\times H$.
\end{example}

Let us consider the tangent morphism
\mar{1aa}\beq
TR_{GP}: (G\times \cG_l)\op\times_X TP \ar_X TP  \label{1aa}
\eeq
to the right action $R_{GP}$ (\ref{1}) of $G$ on $P$. Its
restriction to $T_\bb G\times_X TP$ provides a homomorphism
\mar{52f44}\beq
\cG_l\ni \e\to \xi_\e\in \cT(P) \label{52f44}
\eeq
of the left Lie algebra $\cG_l$ of $G$ to the Lie algebra $\cT(P)$
of vector fields on $P$. Vector fields $\xi_\e$ (\ref{52f44}) are
obviously  vertical. They are called fundamental vector fields
\cite{kob}. Given a basis $\{\e_r\}$ for $\cG_l$, the
corresponding fundamental vector fields $\xi_r=\xi_{\e_r}$ form a
family of nowhere vanishing and linearly independent sections of
the vertical tangent bundle $VP$ of $P\to X$. Consequently, this
bundle is trivial
\mar{ttt91}\beq
VP = P\times\cG_l. \label{ttt91}
\eeq

Restricting the tangent morphism $TR_{GP}$ (\ref{1aa}) to
\mar{2aa}\beq
TR_{GP}: \wh 0(G)\op\times_X TP \ar_X TP,  \label{2aa}
\eeq
we obtain the tangent prolongation of the structure group action
$R_{GP}$ (\ref{1}). Since the action of $G$ (\ref{1}) on $P$ is
fibrewise, its action (\ref{2aa})  is restricted to the vertical
tangent bundle $VP$ of $P$.

Taking the quotient of the tangent bundle $TP\to P$ and the
vertical tangent bundle $VP$ of $P$ by $G$ (\ref{2aa}), we obtain
the vector bundles
\mar{b1.205}\beq
 T_GP=TP/G,\qquad  V_GP=VP/G \label{b1.205}
\eeq
over $X$.  Sections of $T_GP\to X$ are $G$-invariant vector fields
on $P$. Accordingly, sections of $V_GP\to X$ are $G$-invariant
vertical vector fields on $P$. Hence, a typical fibre of $V_GP\to
X$ is the right Lie algebra $\cG_r$ of $G$ subject to the adjoint
representation of a structure group $G$. Therefore, $V_GP$
(\ref{b1.205}) is called the Lie algebra bundle.

Given a bundle atlas $\Psi_P$ (\ref{51f2}) of $P$, there is the
corresponding atlas
\mar{52f57}\beq
\Psi=\{(U_\al,\psi_\al), {\rm Ad}_{\vr_{\al\bt}}\} \label{52f57}
\eeq
of the Lie algebra bundle $V_GP$, which is provided with bundle
coordinates $(U_\al; x^\m,\chi^m)$ with respect to the fibre
frames $\{e_m=\psi_\al^{-1}(x)(\ve_m)\}$, where $\{\ve_m\}$ is a
basis for the Lie algebra $\cG_r$. These coordinates obey the
transformation rule
\be
\vr(\chi^m) \ve_m=\chi^m {\rm Ad}_{\vr^{-1}}(\ve_m).
\ee
A glance at this transformation rule shows that $V_GP$ is a bundle
with a structure group $G$. Moreover, it is associated with a
principal $G$-bundle $P$.

Accordingly, the vector bundle $T_GP$ (\ref{b1.205}) is endowed
with bundle coordinates $(x^\m,\dot x^\m,\chi^m)$ with respect to
the fibre frames $\{\dr_\m,e_m\}$. Their transformation rule is
\be
\vr(\chi^m)\ve_m=\chi^m{\rm Ad}_{\vr^{-1}}(\ve_m) +\dot x^\m
R^m_\m\ve_m.
\ee

The Lie bracket of $G$-invariant vertical vector fields on $P$
goes to the quotient by $G$ and defines the Lie bracket of
sections of the vector bundle $T_VP\to X$. This bracket reads
\mar{1129'}\beq
[\xi,\eta]= c_{pq}^r\xi^p\eta^q e_r. \label{1129'}
\eeq
A glance at the expression (\ref{1129'}) shows that sections of
$V_GP$ form a finite-dimensional Lie $C^\infty(X)$-algebra, called
the gauge algebra.

In classical gauge theory, gauge fields are conventionally
described as principal connections on principal bundles. Principal
connections on a principal bundle $P$ (\ref{51f1}) are connections
on $P$ which are equivariant with respect to the right action
(\ref{1}) of a structure group $G$ on $P$. In order to describe
them, we follow the definition of connections on a fibre bundle
$Y\to X$ as global sections of the affine jet bundle $J^1Y\to Y$
\cite{book09,book00,sard13}.

Let $J^1P$ be the first order jet manifold of a principal
$G$-bundle $P\to X$ (\ref{51f1}). Then connections on a principal
bundle $P\to X$ are global sections
\mar{c4c}\beq
A: P\to J^1P \label{c4c}
\eeq
of the affine jet bundle $J^1P\to P$. In order to define principal
connections on $P\to X$, let us consider the jet prolongation
\be
J^1R_{GP}: J^1(X\times G)\op\times_X J^1P \to J^1P
\ee
of the morphism $R_{GP}$ (\ref{1}). Restricting this morphism to
\be
J^1R_G: \wh 0(G)\op\times_X J^1P \to J^1P,
\ee
we obtain the  jet prolongation of the structure group action
$R_{GP}$ (\ref{1}). It reads
\mar{53f1}\beq
G\ni g: j^1_xp\to (j^1_xp)g =j^1_x(pg). \label{53f1}
\eeq
Taking the quotient of the affine jet bundle $J^1P$ by $G$
(\ref{53f1}), we obtain the affine bundle
\mar{B1}\beq
C=J^1P/G\to X\label{B1}
\eeq
modelled over the vector bundle
\be
\ol C=T^*X\op\ot_X V_GP\to X.
\ee
Hence, there is the vertical splitting $VC= C\ot_X \ol C$ of the
vertical tangent bundle $VC$ of $C\to X$.

\begin{remark} \mar{53r1} \label{53r1} A glance at the expression
(\ref{53f1}) shows that the fibre bundle $J^1P\to C$ is a
principal bundle with the structure group $G$. It is canonically
isomorphic to the pull-back
\mar{b1.251}\beq
J^1P= P_C=C\op\times_X P\to C. \label{b1.251}
\eeq
\end{remark}

Since a connection $A$ (\ref{c4c}) on a principal bundle $P\to X$
is principal connection if it is equivariant under the action
(\ref{53f1}) of a structure group $G$, there is obvious one-to-one
correspondence between the principal connections on a principal
$G$-bundle $P$ and global sections
\mar{BB1}\beq
A:X\to C \label{BB1}
\eeq
of the factor bundle $C\to X$ (\ref{B1}), called the bundle of
principal connections. Since the bundle $C\to X$ is affine,
principal connections on a principal bundle always exist.

A bundle of principal connections $C$ is provided with bundle
coordinates $(x^\la,a^m_\m)$ possessing a transformation rule
\mar{53f10}\beq
\vr(a^m_\m)\ve_m=(a^m_\nu{\rm Ad}_{\vr^{-1}}(\ve_m) +
R^m_\nu\ve_m)\frac{\dr x^\nu}{\dr x'^\m}. \label{53f10}
\eeq
Herewith, there is the canonical bundle monomorphism
\be
\la_C: C\ni (x^\la,a^m_\m)\ar_X dx^\m\ot(\dr_\m + \chi_\m^m
e_m)\in T^*X\op\ot_X T_GP.
\ee
Due to this bundle monomorphism, any principal connection $A$
(\ref{BB1}) is represented by a $T_GP$-valued form
\mar{1131}\beq
A=dx^\la\ot (\dr_\la + A_\la^q e_q)  \label{1131}
\eeq
which provides a splitting of this exact sequence
\be
0\to V_GP\ar_X T_GP\ar TX\to 0.
\ee

In classical gauge theory, global sections $A$ (\ref{BB1}) of the
bundle $C\to X$ of principal connections are treated as gauge
fields \cite{book09,book00,sard13}.

Let us consider first order Lagrangian theory of principal
connections. Its configuration space is the first order jet
manifold $J^1C$ of the bundle of principal connections $C$
(\ref{B1}), endowed with bundle coordinates $(x^\m,a^m_\m)$
possessing transition functions (\ref{53f10}). This configuration
space admits the canonical splitting
\mar{296,'}\ben
&& J^1C =C_+\op\oplus_C C_-=C_+\op\oplus_C (C\op\times_X\op\w^2T^*X\ot V_GP),
\label{296}\\
&&a_{\la\m}^r = \frac12(\cS_{\la\m}^r + \cF_{\la\m}^r)= \frac{1}{2}(a_{\la\m}^r + a_{\m\la}^r
 - c_{pq}^r a_\la^p a_\m^q) + \frac{1}{2}
(a_{\la\m}^r - a_{\m\la}^r + c_{pq}^r a_\la^p a_\m^q).
\label{296'}
\een
Given a section $A$ of a bundle $C\to X$ and its jet prolongation
$j^1A$, the projection of $j^1A$ to $C_-$ is the strength
\mar{1136b}\ben
&& F_A
=\frac12 F^r_{\la\m} dx^\la\w dx^\m\ot e_r, \nonumber \\
&& F_{\la\m}^r = \cF_{\la\m}^r \circ j^1A=\dr_\la A_\m^r - \dr_\m A_\la^r + c_{pq}^rA_\la^p
A_\m^q, \label{1136b}
\een
of a gauge field $A$.

Given a first order Lagrangian
\mar{57f1}\beq
L=\cL\om: J^1C\to \op\w^n T^*X \label{57f1}
\eeq
on $J^1C$, the corresponding Euler--Lagrange operator
(\ref{0709'}) reads
\mar{57f2}\beq
\cE_L= \cE_r^\m\thh^r_\m\w\om=(\dr_r^\m- d_\la\dr^{\la\m}_r)\cL
\thh^r_\m\w\om. \label{57f2}
\eeq
Its kernel defines the Euler--Lagrange equation
\be
\cE_r^\m=(\dr_r^\m- d_\la\dr^{\la\m}_r)\cL=0.
\ee

In Yang--Mills gauge theory, the Lagrangian $L$ (\ref{57f1}) is
assumed to be invariant under principal gauge transformations.
These are vertical principal automorphisms $\Phi_P$ of a principal
$G$-bundle $P$ which are equivariant  under the right action
(\ref{1}) of a structure group $G$ on $P$, i.e.,
\mar{55ff1}\beq
\Phi_P(pg)=\Phi_P(p)g, \qquad g\in G, \qquad p\in P. \label{55ff1}
\eeq

Every vertical principal automorphism of a principal bundle $P$ is
represented as $\Phi_P(p)=pf(p)$, $p\in P$, where $f$ is a
$G$-valued  equivariant function on $P$, i.e.,
\mar{b3115}\beq
f(pg)=g^{-1}f(p)g, \qquad g\in G. \label{b3115}
\eeq
There is one-to-one correspondence between the equivariant
functions $f$ (\ref{b3115}) and the global sections $s$ of the
$P$-associated group bundle $P^G\to X$ whose typical fibre is the
group $G$ which acts on itself by the adjoint representation. This
correspondence is defined by the relation
\be
s(\pi_P(p))p = pf(p), \qquad p\in P.
\ee
The group of vertical principal automorphisms of a principal
$G$-bundle is called the gauge group. It is isomorphic to the
group $P^G(X)$ of global sections of the group bundle $P^G$.

In order to describe principal gauge transformations of the
Lagrangian $L$ (\ref{57f1}), let us restrict our consideration to
(local) one-parameter groups of vertical principal automorphisms
of $P$. Their infinitesimal generators are $G$-invariant vertical
vector fields $\xi$ on $P$. They are represented by sections
\mar{b3106}\beq
\xi=\xi^p e_p \label{b3106}
\eeq
of the Lie algebra bundle $V_GP\to X$ (\ref{b1.205}). We call
$\xi$ (\ref{b3106}) the infinitesimal principal gauge
transformations. They form a Lie $C^\infty(X)$-algebra.

Any (local) one-parameter group of principal automorphism $\Phi_P$
(\ref{55ff1}) of a principal bundle $P$ admits the jet
prolongation $J^1\Phi_P$ to a one-parameter group of
$G$-equivariant automorphism of the jet manifold $J^1P$ which, in
turn, yields a one-parameter group of principal automorphisms
$\Phi_C$ of the bundle of principal connections $C$ (\ref{B1}).
Its infinitesimal generator is a vector field on $C$, called the
principal vector field on $C$ and regarded as an infinitesimal
gauge transformation of $C$. Thus, any infinitesimal principal
gauge transformation $\xi$ (\ref{b3106}) on $P$ yields a vertical
vector field
\mar{279}\beq
u_\xi = (\dr_\m\xi^r + c_{pq}^r a_\m^p\xi^q)\dr_r^\m \label{279}
\eeq
on $C$. Its jet prolongation onto $J^1C$ reads
\be
J^1u_\xi =u_\xi +(\dr_{\la\m}\xi^r + c_{pq}^r a_\m^p\dr_\la\xi^q
+c_{pq}^r a_{\la\m}^p\xi^q)\dr_ r^{\la\m}.
\ee

It is readily justified that the monomorphism
\mar{55f5}\beq
V_GP(X)\ni\xi \to u_\xi\in \cT(C) \label{55f5}
\eeq
obeys the equality
\mar{55f5'}\beq
u_{[\xi,\eta]}=[u_\xi, u_\eta], \label{55f5'}
\eeq
i.e., it is a monomorphism of real Lie algebras. Moreover, a
glance at the expression (\ref{279}) shows that the monomorphism
(\ref{55f5}) is a linear first order differential operator which
sends sections of the pull-back bundle $C\op\times_X V_GP\to C$
onto sections of the vertical tangent bundle $VC\to C$. We
therefore can treat vector fields (\ref{279}) as infinitesimal
gauge transformations depending on gauge parameters $\xi\in
V_GP(X)$ \cite{jmp09,book09}.

Let us assume that the gauge theory Lagrangian $L$ (\ref{57f1}) on
$J^1C$ is invariant under vertical gauge transformations. This
means that vertical vector fields (\ref{279}) are exact symmetries
of $L$, i.e.,
\mar{57f4}\beq
\bL_{J^1u_\xi}L=0. \label{57f4}
\eeq
 Then  the first variational formula (\ref{g107}) for the Lie
derivative (\ref{57f4}) takes a form
\mar{57f5}\beq
0= (\dr_\m\xi^r + c_{pq}^r a_\m^p\xi^q)\cE_r^\m +
d_\la[(\dr_\m\xi^r + c_{pq}^r a_\m^p\xi^q)\dr^{\la\m}_r\cL)].
\label{57f5}
\eeq
It leads to the gauge invariance conditions
\mar{57f7a-c}\ben
&& \dr_p^{\m\la}\cL + \dr_p^{\la\m}\cL = 0, \label{57f7a}\\
&& \cE_r^\m +d_\la \dr^{\la\m}_r\cL
+ c_{pr}^q a_\nu^p\dr^{\m\nu}_q\cL=0, \label{57f7b}\\
&& c_{pq}^r(a_\m^p \cE^\m_r+
d_\la( a_\m^p\dr^{\la\m}_r\cL))=0. \label{57f7c}
\een

One can regard the equalities (\ref{57f7a}) -- (\ref{57f7c}) as
the conditions of a Lagrangian $L$ to be gauge invariant. They are
brought into the form
\mar{57f8a-c}\ben
&& \dr_p^{\m\la}\cL + \dr_p^{\la\m}\cL = 0. \label{57f8a}\\
&& \dr_q^\m \cL + c_{pq}^r a_\nu^p \dr_r^{\m\nu}\cL  = 0, \label{57f8b} \\
&& c_{pq}^r(a_\m^p\dr_r^\m\cL + a^p_{\la\m} \dr_r^{\la\m}\cL)  =
0. \label{57f8c}
\een
Let us utilize the coordinates $(a^q_\m, \cF^r_{\la\m},
\cS^r_{\la\m})$ (\ref{296'}) which correspond to the canonical
splitting (\ref{296}) of the affine jet bundle $J^1C\to C$. With
respect to these coordinates, the equation (\ref{57f8a}) reads
\beq
\frac{\dr\cL}{\dr {\cal S}^p_{\m\la}}=0. \label{b3128}
\eeq
Then the equation (\ref{57f8b}) takes a form
\beq
\frac{\dr\cL}{\dr a^q_\m}=0. \label{b3129}
\eeq
A glance at the equalities (\ref{b3128}) and (\ref{b3129}) shows
that a gauge invariant Lagrangian factorizes through the strength
coordinates $\cF$ (\ref{296'}). Then the equation (\ref{57f8c}),
written as
\be
c^r_{pq}\cF^p_{\la\m}\frac{\dr\cL}{\dr \cF^r_{\la\m}}=0,
\ee
shows that the gauge symmetry $u_\xi$ of a Lagrangian $L$ is
exact. The following thus has been proved.

\begin{theorem} \mar{57t1} \label{57t1} A gauge theory Lagrangian
(\ref{57f1}) possesses the exact gauge symmetry $u_\xi$
(\ref{279}) only if it factorizes through the strength $\cF$
(\ref{296'}).
\end{theorem}

A corollary of this result is the well-known Utiyama theorem
\cite{bruz}.

\begin{theorem} There is a unique gauge invariant quadratic first order
Lagrangian (with the accuracy to variationally trivial ones). It
is the conventional Yang--Mills Lagrangian
\mar{5.1}\beq
L_{\rm YM}=\frac14a^G_{pq}g^{\la\m}g^{\bt\n}\cF^p_{\la
\beta}\cF^q_{\m\n}\sqrt{|g|}\,\om, \qquad  g=\det(g_{\m\nu}), \label{5.1}
\eeq
where $a^G$ is a $G$-invariant bilinear form on the Lie algebra
$\cG_r$ and $g$ is a world metric on $X$.
\end{theorem}

The Euler--Lagrange operator (\ref{57f2}) of the Yang--Mills
Lagrangian $L_{\rm YM}$ (\ref{5.1}) is
\mar{57f13}\beq
\cE_{\rm YM}=\cE^\m_r\thh^\m_r\w\om=(\dl^n_rd_\la
+c^n_{rp}a^p_\la)(a^G_{nq}g^{\m\al}g^{\la\bt}
\cF^q_{\al\bt}\sqrt{|g|})\thh_\m^r\w\om. \label{57f13}
\eeq
Its kernel defines the Yang--Mills equations
\be
\cE^\m_r= (\dl^n_rd_\la
+c^n_{rp}a^p_\la)(a^G_{nq}g^{\m\al}g^{\la\bt}
\cF^q_{\al\bt}\sqrt{|g|})=0.
\ee

We call a Lagrangian system $(\cS^*_\infty[C], L_{\rm YM})$ the
Yang--Mills gauge theory.

\begin{remark} In classical gauge theory, there are Lagrangians, e.g., the
Chern--Simons one which do not factorize through the strength of a
gauge field, and whose gauge symmetry $u_\xi$ (\ref{279}) is
variational, but not exact.
\end{remark}

Since the gauge symmetry $u_\xi$ of the Yang--Mills Lagrangian
(\ref{5.1}) is exact, the first variational formula (\ref{57f5})
leads to the weak conservation law (\ref{35f2}):
\mar{57f10}\beq
0\ap d_\la(-u_\xi{}^\m_r\dr^{\la\m}_r\cL_{\rm YM}), \label{57f10}
\eeq
of the Noether current (\ref{42f50}):
\mar{57f11}\beq
\cJ^\la_\xi=-(\dr_\m\xi^r + c_{pq}^r a_\m^p\xi^q)
(a^G_{rq}g^{\m\al}g^{\la\bt} \cF^q_{\al\bt}\sqrt{|g|}).
\label{57f11}
\eeq
In accordance with Theorem \ref{supp'}, the Noether current
(\ref{57f11}) is brought into the superpotential form
(\ref{b381'}) which reads
\be
\cJ^\la_\xi= \xi^r\cE_r^\m + d_\nu(\xi^r\dr_r^{[\nu\m]}\cL_{\rm
YM}), \qquad U^{\nu\m}= \xi^r a^G_{rq}g^{\nu\al}g^{\m\bt}
\cF^q_{\al\bt}\sqrt{|g|}.
\ee

The gauge invariance conditions (\ref{57f7a}) -- (\ref{57f7c})
lead to the Noether identities
\mar{57f6}\beq
c^p_{rq}a^q_\m\cE_p^\m + d_\m\cE_r^\m=0. \label{57f6}
\eeq
which the Euler--Lagrange operator $\cE_{\rm YM}$ (\ref{57f13}) of
the Yang--Mills Lagrangian (\ref{5.1}) satisfies. These Noether
identities are associated to the gauge symmetry $u_\xi$
(\ref{279}). One can show the following \cite{book09}.

\begin{lemma} \mar{57l1} \label{57l1} The Noether identities (\ref{57f6}) are
non-trivial, complete an irreducible..
\end{lemma}

Thus, Yang--Mills gauge theory is an irreducible degenerate
Lagrangian theory characterized by the complete Noether identities
(\ref{57f6}).

Following inverse second Noether Theorem \ref{w35}, let us
consider the differential bigraded algebra
\mar{vvc1}\beq
P^*_\infty\{0\}=\cS^*_\infty[\ol{VC}\op\oplus_C
V_GP;C\op\times_X\ol {V_GP}] \label{vvc1}
\eeq
with the local generating basis $(a^r_\m,\ol a^\m_r, c^r, \ol
c_r)$ where $c_r$ are odd ghosts. The gauge operator $u$
(\ref{w33}) associated to the Noether identities (\ref{57f6})
reads
\mar{57f30}\beq
u=(c_\m^r + c_{pq}^r a_\m^pc^q)\dr_r^\m. \label{57f30}
\eeq
It is an odd gauge symmetry of the Yang--Mills Lagrangian $L_{\rm
YM}$ which can be obtained from the gauge symmetry $u_\xi$
(\ref{279}) by replacement of gauge parameters $\xi^r$ with odd
ghosts $c^r$.

Furthermore, due to the relation (\ref{55f5'}), the gauge operator
$u$ (\ref{57f30}) admits the nilpotent BRST extension
\be
\bbc= (c_\m^r + c_{pq}^r a_\m^pc^q)\frac{\dr}{\dr a_\m^r} -\frac12
c^r_{pq}c^pc^q\frac{\dr}{\dr c^r},
\ee
which is the well-known BRST operator in Yang--Mills gauge theory
\cite{gom}. Then, by virtue of Theorem \ref{w130}, the Yang--Mills
Lagrangian $L_{\rm YM}$ is extended to a proper solution of the
master equation
\be
L_E=L_{\rm YM}+ (c_\m^r + c_{pq}^r a_\m^pc^q)\ol a^\m_r\om
-\frac12 c^r_{pq}c^pc^q\ol c_r\om.
\ee

\section{Principal graded bundles}

By analogy to gauge theory on principal bundles, we develop SUSY
gauge theory as theory of graded principal connections on
principal graded bundles.

Graded principal bundles and connections on these bundles
\cite{book00,sard09,stavr} can be studied similarly to principal
superbundles and principal superconnections
\cite{bart,book00,sard09}. Since the definition of a graded Lie
group and its Lie superalgebra involves the notion of a Hopf
algebra, we start with the following construction \cite{stavr}.

Let $(Z,\gA)$ be a graded manifold of dimension $(n,m)$. One
considers the finite dual $\gA(Z)^\circ$ of its structure ring
$\gA(Z)$ which consists of elements $a$ of the dual $\gA(Z)^*$ of
$\gA(Z)$ vanishing on an ideal of $\gA(Z)$ of finite codimension.
This is brought into a graded commutative coalgebra with the
comultiplication
\be
\Delta^\circ(a)(f\ot f')=a(ff'), \qquad  f,f'\in \gA(Z), \qquad
a\in \gA(Z)^\circ,
\ee
and the counit
\be
\e^\circ(a)= a(\bb_{\gA(Z)}).
\ee
In particular, $\gA(Z)^\circ$ includes the evaluation elements
$\dl_z$ such that
\be
\dl_z(f)=(\si(f))(z).
\ee
Given an evaluation element $\dl_z$, elements $u\in\gA(Z)^\circ$
are said to be primitive relative to $\dl_z$ if they obey a
relation
\be
\Delta^\circ(u)= u\ot\dl_z +\dl_z\ot u.
\ee
These elements are derivations of $\gA(Z)$ at $z$, i.e.,
\be
u(ff')=u(f)\dl_z(f') +(-1)^{[u][f]}\dl_z(f)u(f').
\ee

\begin{definition} \label{su70} \mar{su70}
A graded Lie group $(G,\ccG)$ is defined as a graded manifold
whose body $G$ is a real Lie group, and the structure ring
$\ccG(G)$ is a graded Hopf algebra $(\Delta, \e, S)$ such that a
body epimorphism $\si:\ccG(G)\to C^\infty(G)$ is a Hopf algebra
morphism where a ring $C^\infty(G)$ of smooth real functions on a
Lie group $G$  possesses a real Hopf algebra structure with the
co-operations:
\be
\Delta(f)(g,g')=f(gg'), \qquad \e(f)=f(\bb), \qquad
S(f)(g)=f(g^{-1}), \qquad f\in C^\infty(G).
\ee
\end{definition}

One can show that the finite dual $\ccG(G)^\circ$ of
$\ccG(G)^\circ$ is equipped with the structure of a real Hopf
algebra with the multiplication law
\mar{+236}\beq
a\star b=(a\ot b)\circ\Delta, \qquad  a,b\in\ccG(G)^\circ.
\label{+236}
\eeq
With respect to this multiplication, the evaluation elements
$\dl_g$, $g\in G$, provided with the product
$\dl_g\star\dl_{g'}=\dl_{gg'}$ (\ref{+236}), constitute a group
isomorphic to $G$. They are called the group-like elements.  It is
readily observed that the set of primitive elements of
$\ccG(G)^\circ$ relative to $\dl_\bb$ is a real Lie superalgebra
${\mathfrak g}$ with respect to the bracket
\mar{su71}\beq
[u,u']=u\star u' -(-1)^{[u][u']}u'\star u. \label{su71}
\eeq
It is called the Lie superalgebra of a graded Lie group
$(G,\ccG)$. Its even part $\cG_0$ is a Lie algebra of a Lie group
$G$.

One says that a graded Lie group $(G,\ccG)$ acts on a graded
manifold $(Z,\gA)$ on the right if there exists a morphism of
graded manifolds
\be
(\vf,\Phi):(Z,\gA)\times(G,\ccG)\to (Z,\gA)
\ee
such that the corresponding ring morphism
\be
\Phi:\gA(Z)\to \gA(Z)\ot \ccG(G)
\ee
defines a structure of a right $\ccG(G)$-comodule on $\gA(Z)$,
i.e.,
\be
(\id\ot \Delta)\circ\Phi=(\Phi\ot\id)\circ\Phi, \qquad
(\id\ot\e)\circ\Phi=\id.
\ee
For a right action $(\vf,\Phi)$ and for each element
$a\in\ccG(G)^\circ$, one can introduce a linear map
\mar{+235}\beq
\Phi_a=(\id\ot a)\circ\Phi: \gA(Z)\to\gA(Z). \label{+235}
\eeq
In particular, if $a$ is a primitive element with respect to
$\dl_e$, then $\Phi_a\in\gd \gA(Z)$.

Let us consider a right action of a graded Lie group $(G,\ccG)$ on
itself. If $\Phi=\Delta$ and $a=\dl_g$ is a group-like element,
then
\mar{su90}\beq
r_g=(\id\ot \dl_g)\circ\Delta \label{su90}
\eeq
is an even graded algebra isomorphism which  corresponds to the
right translation $G\to Gg$. Similarly, the left action
\mar{su91}\beq
l_g=(\dl_g\ot\id)\circ\Delta \label{su91}
\eeq
is defined. If $a\in{\mathfrak g}$, then $l_a$ is a derivation of
$\ccG(G)$. Given a basis $\{u_i\}$ for ${\mathfrak g}$, the
derivations $\Phi_{u_i}$  constitute the global basis for
$\gd\ccG(G)$, i.e., $\gd\ccG(G)$ is a free left $\ccG(G)$-module.
In particular, there is a decomposition
\be
&&\ccG(G)=\ccG'(G)\oplus_R\ccG''(G),\\
&& \ccG'(G)=\{ f\in\ccG(G)\, :\, \Phi_u(f)=0, \,\,  u\in {\mathfrak
g}_0\},\\
&& \ccG''(G)=\{ f\in\ccG(G)\, :\, \Phi_u(f)=0, \,\,  u\in {\mathfrak
g}_1\}.
\ee
Since $\ccG''(G)\cong C^\infty(G)$, one can show the following
\cite{alm,boy,kost77}.

\begin{theorem} \label{su100} \mar{su100}
A graded Lie group $(G,\ccG)$ is a simple graded manifold modelled
over a trivial bundle
\mar{su101}\beq
G\times \w{\mathfrak g}_1^*\to G, \label{su101}
\eeq
endowed with the right action $r_g$, $g\in G$, (\ref{su90}) of a
group $G$.
\end{theorem}

Let us turn now to the notion of a principal graded bundle, but,
following Section 5, we restrict our consideration to graded
bundles over a smooth manifold $X$. A right action $(\vf,\Phi)$ of
graded Lie group $(G,\ccG)$ on a graded manifold $(Z,\gA)$ is
called free if, for each $z\in Z$, the morphism $\Phi_z:\gA(Z)\to
\ccG(G)$ is such that the dual morphism $\Phi_{z*}:\ccG(G)^\circ
\to \gA(Z)^\circ$ is injective. A right action $(\vf,\Phi)$ of
$(G,\ccG)$ on $(Z,\gA)$ is said to be regular if the morphism
\be
(\vf\times\pr_1)\circ\Delta: (Z,\gA)\times (G,\ccG)\to
(Z,\gA)\times (Z,\gA)
\ee
defines a closed graded submanifold of $(Z,\gA)\times (Z,\gA)$.

\begin{remark}
Let us note that $(Z',\gA')$ is said to be a graded submanifold of
$(Z,\gA)$ if there exists a morphism $(Z',\gA')\to (Z,\gA)$ such
that the corresponding morphism $\gA'(Z')^\circ\to \gA(Z)^\circ$
is an inclusion. A graded submanifold is called closed if ${\rm
dim}\,(Z',\gA')<{\rm dim}\,(Z,\gA)$.
\end{remark}

Then we come to the following variant of the well-known theorem on
the quotient of a graded manifold \cite{alm,stavr}.

\begin{theorem}
A right action $(\vf,\Phi)$ of $(G,\ccG)$ on $(Z,\gA)$ is regular
iff the quotient $(Z/G,\gA/\ccG)$ is a graded manifold, i.e.,
there exists an epimorphism of graded manifolds $(Z,\gA)\to
(Z/G,\gA/\ccG)$ compatible with the surjection $Z\to Z/G$.
\end{theorem}

In view of this Theorem,  a principal graded bundle $(P,\gA)$ can
be defined as a locally trivial submersion
\be
(P,\gA)\to (P/G,\gA/\ccG)
\ee
with respect to the right regular free action of $(G,\ccG)$ on
$(P,\gA)$.  In an equivalent way, one can say that a principal
graded bundle is a graded manifold  $(P,\gA)$ together with a free
right action of a graded Lie group $(G,\ccG)$ on  $(P,\gA)$ such
that the quotient $(P/G,\gA/\ccG)$ is a graded manifold and the
natural surjection $(P,\gA)\to (P/G,\gA/\ccG)$ is a submersion.
Obviously, $P\to P/G$ is a familiar principal bundle with a
structure group $G$.

As was mentioned above, we restrict our consideration to the case
of a principal graded bundle whose base a trivial graded manifold
$(X=P/G,\gA/\ccG=C^\infty_X)$. i.e., principal graded bundles over
a smooth manifold $X$.

\begin{definition} \label{su55} \mar{su55} A principal graded
bundle is defined to be is a simple graded manifold $(P, \gA_F)$
whose body is a principal bundle $P\to X$ with a structure group
$G$, and which is modelled over a composite bundle
\mar{su52}\beq
F=P\op\times_X W\to P\to X \label{su52}
\eeq
where $W\to X$ is $P$-associated bundle with a typical fibre
$\cG_1$ provided with the left action (\ref{su91}) of a group $G$.
A principal graded bundle $(P, \gA_F)$ is subject to the right
fibrewise action (\ref{+235}) of a group $G$.
\end{definition}

\begin{remark} \label{su92} \mar{su92}
Being kept as a simple graded manifold, a graded bundle $(P,
\gA_F)$ fails to admit an action of a graded Lie group $(G,\ccG)$.
Therefore, it is not a principal graded bundle in a strict sense.
\end{remark}

\section{Yang--Mills supergauge theory}

Let $\cG$ be a finite-dimensional real Lie superalgebra with the
basis $\{\ve_r\}$, $r=1,\ldots,m,$ and real structure constants
$c^r_{ij}$ such that
\be
&& c^r_{ij}=-(-1)^{[i][j]}c^r_{ji}, \qquad [r]=[i]+[j],\\
&& (-1)^{[i][b]}c^r_{ij}c^j_{ab} + (-1)^{[a][i]}c^r_{aj}c^j_{bi} +
(-1)^{[b][a]}c^r_{bj}c^j_{ia}=0,
\ee
where $[r]$ denotes the Grassmann parity of $\ve_r$. Its even part
$\cG_0$ is a Lie algebra of some Lie group $G$. Let the adjoint
representation of $\cG_0$ in $\cG$ is extended to the
corresponding action of $G$ on $\cG$.

Let $P$ be a principal bundle with a structure group $G$. Given
Yang--Mills gauge theory of principal connections on $P$ (see
Section 9), we aim to extend it to a Yang--Mills supergauge theory
associated to a Lie superalgebra $\cG$.

Let us consider a simple graded manifold $(G,\ccG)$ modelled over
the trivial vector bundle (\ref{su101}). In accordance with
Theorem \ref{su100}, this is a graded Lie group whose Lie
superalgebra is $\cG$. Following Definition \ref{su55}, let us
define a principal graded bundle $(P,\gA_F)$ modelled over the
composite bundle (\ref{su52}).

Let $J^1F$ be a jet manifold  of the fibre bundle $F\to X$
(\ref{su52}). It is a composite bundle
\mar{su61}\beq
J^1F\to J^1P\to X, \label{su61}
\eeq
where $J^1F\to J^1P$ is a vector bundle. Then a simple graded
manifold $(J^1P, A_{J^1f})$ modelled over $J^1F\to J^1P$ is a
graded jet manifold of a principal graded bundle $(P, \gA_F)$ in
accordance with Definition \ref{su32}. The composite bundle
(\ref{su61}) is provided with the jet prolongation of the right
action (\ref{+235}) of $G$ onto $(P, \gA_F)$. By analogy with the
bundle of principal connections $C$ (\ref{B1}), let us consider
the quotient
\mar{su60}\beq
\sC=J^1F/G\to C\to X. \label{su60}
\eeq
Since $\sC\to C$ is a vector bundle, we have a simple graded
manifold $(C,\gA_\sC)$ modelled over a vector bundle $\sC\to C$.
It is a graded bundle over a smooth manifold $X$.

We develop supergauge theory as a Lagrangian theory on a graded
bundle $(C,\gA_\sC)$ (see Section 7). Given a basis $\{\ve_r\}$
for $\cG$, we have a local basis $(x^\la, a^r_\la)$  of Grassmann
parity $[a^r_\la]=[r]$ for a graded bundle $(C,\gA_\sC)$.
Similarly to the splitting (\ref{296'}), jets of the elements
$a^r_\la$ admit the decomposition
\mar{f31}\ben
&& a^r_{\la\m}=\frac12(\cF^r_{\la\m} +
\cS^r_{\la\m})= \label{f31}\\
&& \qquad \frac12(a^r_{\la\m}-a^r_{\m\la} +c^r_{ij}a^i_\la a^j_\m)
+\frac12(a^r_{\la\m}+ a^r_{\m\la} -c^r_{ij}a^i_\la
a^j_\m).\nonumber
\een

Let us consider the differential bigraded algebra
$\cS^*_\infty[\sC;C]$ (\ref{5.77a}). Given the universal
enveloping algebra $\ol \cG$ of $\cG$, we assume that there exists
an even quadratic Casimir element $h^{ij}\ve_i\ve_j$ of $\ol\cG$
such that the matrix $h^{ij}$ is non-degenerate. Let $g$ be metric
on $X$. Then a graded Yang--Mills Lagrangian is defined as
\be
L_{\rm YM}=\frac14
h_{ij}g^{\la\m}g^{\bt\nu}\cF^i_{\la\bt}\cF^j_{\m\nu}\om.
\ee
Its variational derivatives $\cE_r^\la$ obey the irreducible
complete Noether identities
\be
c^r_{ji}a^i_\la\cE_r^\la +d_\la\cE_j^\la=0.
\ee

Therefore, let us enlarge the differential bigraded algebra
$\cS^*_\infty[\sC;C]$ to the differential bigraded algebra
\be
P^*_\infty\{0\}=\cS^*_\infty[\sC\op\oplus_X \ol\sC,\op\oplus_X
VF/G\op\oplus_X \ol{VF/G};P],
\ee
where $\ol E$ denotes the density dual (\ref{41f2}) of a fibre
bundle $E\to X$. Its local basis $(x^\la, a^r_\la,\ol a^\la_r,
c^r, \ol c_r)$ contains gauge fields $a^r_\la$, their antifields
$\ol a^\la_r$ of Grassmann parity
\be
[\ol a_r^\la]= ([r]+1){\rm mod}\,2,
\ee
the ghosts $c^r$ of Grassmann parity
\be
[c^r]= ([r]+1){\rm mod}\,2,
\ee
and the Noether antifields $\ol c_r$ of Grassmann parity $[\ol
c_r]=[r]$. Then the gauge operator (\ref{w33}) reads
\mar{su110}\beq
u= (c^r_\la - c^r_{ji}c^ja^i_\la)\frac{\dr}{\dr a_\la^r}
\label{su110}
\eeq
(cf. (\ref{57f30})). It admits the nilpotent BRST extension
\be
\bbc=(c^r_\la - c^r_{ji}c^ja^i_\la)\frac{\dr}{\dr a_\la^r}
-\frac12 (-1)^{[i]} c^r_{ij}c^ic^j\frac{\dr}{\dr c^r}.
\ee
The corresponding proper solution (\ref{w133}) of the master
equation takes a form
\be
L_E=L_{\rm YM}+ (c^r_\la - c^r_{ji}c^ja^i_\la)\ol a^\la_r\om
-\frac12 (-1)^{[i]} c^r_{ij}c^ic^j\ol c_r\om.
\ee

\end{document}